\documentclass[prodmode]{acmsmall} 

 \usepackage{algorithm}
\usepackage{listings}
\usepackage{color}
\definecolor{gray}{rgb}{0.4,0.4,0.4}
\definecolor{darkblue}{rgb}{0.0,0.0,0.6}
\definecolor{cyan}{rgb}{0.0,0.6,0.6}
\usepackage{multirow}
 
\usepackage{caption}

 \usepackage{algpseudocode}
\usepackage{lipsum}

\usepackage[colorinlistoftodos,prependcaption,textsize=tiny]{todonotes}
 
\setcopyright{none}

\begin{document}


\title{A Systematic and Semi-Automatic Safety-Based Test Case Generation Approach Based on Systems-Theoretic Process Analysis}
\author{Asim Abdulkhaleq
\affil{Institute of Software Technology, University of Stuttgart, Germany}
Stefan Wagner
\affil{Institute of Software Technology, University of Stuttgart, Germany}
}

 
 

\begin{abstract}
   Software safety is a crucial aspect during the development of modern safety-critical systems. Software is becoming responsible for most of the critical functions of systems. Therefore, the software components in the systems need to be tested extensively against their safety requirements to ensure a high level of system safety. However, performing testing exhaustively to test all software behaviours is impossible. Numerous testing approaches exist. However, they do not directly concern the information derived during the safety analysis. STPA (Systems-Theoretic Process Analysis) is a unique safety analysis approach based on system and control theory, and was developed to identify unsafe scenarios of a complex system including software. In this paper, we present a systematic and semi-automatic testing approach based on STPA to generate test cases from the STPA safety analysis results to help software and safety engineers to recognize and reduce the associated software risks. We also provide an open-source safety-based testing tool called \emph{STPA TCGenerator} to support the proposed approach. We illustrate the proposed approach with a prototype of a software of the Adaptive Cruise Control System (ACC) with a stop-and-go function with a Lego-Mindstorms EV3 robot.   
  
\end{abstract}

%
%



%
%




\begin{bottomstuff}

Author's addresses: A. Abdulkhaleq {and}, S. Wagner,  Institute of Software Technology, University of Stuttgart, 
Universit\"atsstraße 38, 70569 Stuttgart, Germany 
\end{bottomstuff}

\maketitle

\section{Introduction}

Software has become an indispensable part of many modern systems and often performs the main safety-critical functions. Hence, software safety must be analysed in a system context to gain a comprehensive understanding of the roles of software and to identify the software-related risks that can cause hazards in the system. Software safety as stated in \cite{NASA1999} is practically concerned with the software causal factors that are linked to individual hazards and ensured that the mitigation of each causal factor is traced from software requirements to design, implementation, and test. A software failure may lead to catastrophic results such as injury or loss of human life, damaged property or environmental disturbances. Therefore, it becomes essential to test the software components for unexpected behaviour before using them in practice \cite{DDS}. The Toyota Prius, the General Motors airbag and the loss of the Mars Polar Lander (MPL) mission \cite{MPL2000} are well-known software problems in which the software played an important role in the loss, although the software had been successfully verified against all functional requirements.

Software testing is a crucial process to assess the quality of the software and determine whether it meets its specified requirements. The term software safety testing  \cite{NASA2004} was introduced and implies that software testing should not only address functional requirements, but the software safety requirements as well. Therefore, the process for testing safety-critical software combines conventional testing and safety analysis approaches to focus the testing efforts in a specific way to address the safety of the software and test the critical risky situations. Fault Tree Analysis (FTA) \cite{FTA1981} and Failure Mode, Effects and Criticality Analysis (FMECA) \cite{FMECA1967} are the approaches commonly used for the purpose of safety-based testing. However, these approaches focus only on single component failures and they have limitations to cope with complex systems including software. Leveson \cite {leveson2011engineering} noted that the primary safety problem in software-intensive systems is not software ``failure" but the lack of appropriate constraints on software behaviour. The solution is to identify the required constraints and enforce them in the software and overall system design. Therefore, a new safety analysis technique called STPA \cite{leveson2011engineering} has been developed to overcome the limitations of the traditional techniques in addressing the unsafe scenarios of complex systems.

\subsection{Problem Statement}

The complexity of safety-critical software makes exhaustive software testing impossible. Therefore, we need to make sure that safety is sufficiently considered. Yet, many existing testing approaches and tools do not incorporate information from safety analysis. In case they do, they rely on traditional safety analysis techniques such as FTA and FMECA which focus on component failures instead of component interactions. A software safety testing approach integrated with alternative systems-theoretic safety analysis approaches such as STPA has been missing. 

\subsection{Research Objective}

Our overall goal is to help software developers to more easily and effectively test safety-critical systems. In this paper, we focus on
filling the aforementioned gap by a method which integrates generating safety-based test cases with the information derived during an STPA safety analysis in a systematic and semi-automatic way.

\subsection{Contributions} 
To reach our research objective, we provide five main contributions: (1) We explore how to apply STPA to safety-critical software. (2) We provide an algorithm based on STPA to derive unsafe software scenarios and automatically translate them into a formal specification in LTL (Linear Temporal Logic) \cite{Pnueli:1977:TLP:1382431.1382534} including timing. (3) We make use of specific STPA features (e.g. process model and the STPA safety requirements) to systematically derive a safe behavioural model. (4) We provide an algorithm to extract a safe test model from the safe behavioural model and check its correctness by automatically transforming it into an SMV representation (Symbolic Model Verifier) \cite{McMillan:1993:SMC:530225} and verify it against the STPA safety requirements using the NuSMV model checker \cite{Cimatti1999}. (5) We propose an algorithm to automatically generate the traceability matrix between the STPA software safety requirements and the test model and generate the safety-based test cases for each safety requirement from the test model.  

\subsection{Terminology}
We define the most relevant terms in table I to ensure a consistent terminology in this paper.
\begin{table*} [h]
	\def\arraystretch{1.1} 
	\renewcommand{\arraystretch}{1.4}
	\tbl{Terminology} {
		\begin{tabular}{ p{3.5cm}     p{9.0cm}   }
			
			Terminology                                                                                                                                    & Definition  
			\\ \hline
			\textbf{Software Safety} & is the discipline of software assurance that is a systematic approach to identifying, analyzing, tracking, mitigating, and controlling software hazards and hazardous functions (data and commands) to ensure safe operation within a system \cite{NASA2004}.\\
			
			\textbf{Accident}        &     Accident (Loss) results from inadequate enforcement of the behavioural safety constraints on the process \cite{leveson2011engineering}.
			\\   
			\textbf{Hazard}     &       Hazard is a system state or set of conditions that, together with a particular set of worst-case environmental conditions, will lead to an accident \cite{leveson2011engineering}.
			\\    	 
			\textbf{Unsafe Control Actions}      &  The hazardous scenarios which might occur in the system due to provided or not provided control action when required  \cite{leveson2011engineering}.  
			\\  
			\textbf{Safety Constraints }    &    The safety constraints are the safeguards which prevent the system from leading to losses (accidents) \cite{leveson2011engineering}.     		\\ 		 
		
			\textbf{Process model }    &  The process model is a model required to determine the environmental and system variables and states that affect the safety of the control actions and it is updated through various forms of feedback.    \cite{leveson2011engineering} \cite{Thomas2013}.       		\\ 	
			
			\textbf{Process model variables }    &  The process model variables are the  safety-critical variables of the controller in the control structure diagram which have an effect on the safety of issuing the control actions  \cite{Thomas2012}.       		\\ 		 
			\textbf{Causal Factors}    &  Causal factors are the accident scenarios that explain how unsafe control actions might occur and how safe control actions might not be followed or executed \cite{leveson2011engineering} \cite{Thomas2012}. 
			\\  
			
				\textbf{ٍSafe Behavioural Model}    &  The safe behavioural model is a statechart notation that models the process model of a software controller in the STPA control structure diagram and and it is constrained by the STPA-generated software safety requirements (transition conditions).
			\\

	\textbf{Safe Test Model} & The safe test model is an extended finite state machine model which is automatically constructed from the safe behavioural model.  \\
	
	\textbf{ٍSafety-based Test Cases}    &  The safety-based test cases are set of the test cases which are generated from information derived during the safety analysis process.
	\\

			\hline
	
	\end{tabular}}
 
	\end{table*}

 \section{Background}

 \subsection{STPA Safety Analysis \& Software Safety}

\begin{figure}[!t]

 	\centering

 	\includegraphics[width=4.5in]{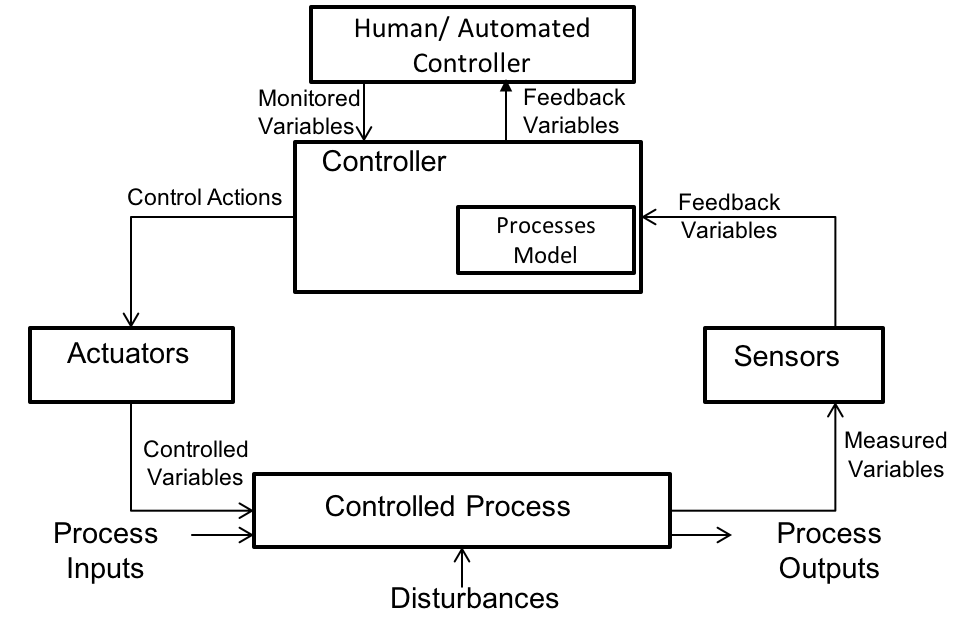}

 	\caption{A general feedback control structure of a software system}

 	\label{fig_sim}

 \end{figure}

 STPA (Systems-Theoretic Processes Analysis) \cite{leveson2011engineering} is a top-down process based on the accident model called STAMP (Systems-Theoretic Accident Model and Processes). STPA is developed for generating detailed safety requirements of complex and modern systems to prevent the occurrence of unsafe scenarios in the systems. In STPA, the system is seen as a set of interrelated components which interact with each other to provide a dynamic equilibrium through feedback loops of information and control. STPA has the following main steps: (1) Establish the fundamentals of the analysis (e.g. system-level accidents and the associated hazards) and draw the control structure diagram of the system (shown in Fig. \ref{fig_sim}). (2) STPA Step 1: Use the control structure diagram to identify the potential unsafe control actions. (3) STPA Step 2: Determine how each potentially unsafe control action could occur by identifying the process model and its variables for each controller and analysing each path in the control structure diagram.

 The basic components in STPA are safety constraints, unsafe control actions, unsafe scenarios, control structure diagram and process models. A control structure diagram is made up of basic feedback control loops. An example is shown in Fig. \ref{fig_sim}. When put together, they can be used to model the high-level control structure of a particular system.

 An extended approach to STPA is proposed by Thomas \cite{Thomas} for identifying the unsafe control actions which are identified in STPA Step 1 based on the combinations of process model variables (context tables) of each controller in the control structure diagram. It also aims at performing STPA STPA 2 based on a set of unsafe control actions which are identified in of STPA Step 1. Thomas \cite{Thomas} mathematically discussed the formalization of STPA which can be used not only to identify unsafe control actions and other control flaws, but also to generate model-based requirements that will enforce safe behaviours.

 \begin{definition} [A Control Structure Diagram] The Control Structure Diagram ($\mathit{CSD}$) of a software system can be expressed with five-tuples (\mbox{$\mathit{CO}$, $\mathit{AC}$, $\mathit{SO}$, $\mathit{CP}$, $\mathit{CA}$}), where $\mathit{CO}$ is a set (one or more) of the software controllers which control the controlled processes ($\mathit{CP}$) by issuing control actions to the actuators, $\mathit{AC}$ is a set of the actuators which implement the control actions ($\mathit{CA}$) of the controller, $\mathit{CP}$ is a set of the controlled processes which are controlled by controllers ($\mathit{CO}$). $\mathit{SO}$ is a set of sensors which send the feedback about the status of the controlled process.  \end{definition}

 Each controller in the control structure diagram must contain a model of the assumed state of the controlled process, called the process model \cite{leveson2011engineering}. A process model contains one or more variables, the required relationships among the variables, the current state and the logic of how the process can change state. This model is used to determine what control actions are needed. It is updated through various forms of feedback \cite{leveson2011engineering}. The process model is a part of the internal state of the controller in the control structure diagram.

 \begin {definition} [A Software Controller] A software controller $\mathit{CO}_{i}$ can be expressed formally as a two-tuple $\mathit{CO}_{i}= (\mathit{CA}, \mathit{PM})$, where $\mathit{CA}$ is set of the control actions and $\mathit{PM}$ is the  process model of the controller which has a set of process model variables ($\mathit{PMV}$), which are a set of critical variables $P$ and states $S$ that have an effect on the safety of $\mathit{CA}$: $\mathit{P}=\bigcup ({\cal P}_{1}=v_{1} \ldots {\cal P}_{n}=v_{n})$, where  $P_{1}$ and $P_{n}$ are process model variables of the software controller $\mathit{CO}_{i}$ with their values $v_{1}$ and $v_{n}$.    \end{definition}

In \cite{Abdulkhaleq20152}, we classified the process model variables of the software controller that affect the safety of the critical control actions into three types: 1)  \emph{Internal variables} which change the status of the software controller,  2) \emph{Interaction interface variables} which receive and store the data/command/feedback from the other components in the system, and 3) \emph{Environmental variables} of the environmental components that interact with or are controlled by the software controller.

\begin{figure*}[t]

\centering

\includegraphics[ width=4.5in ]{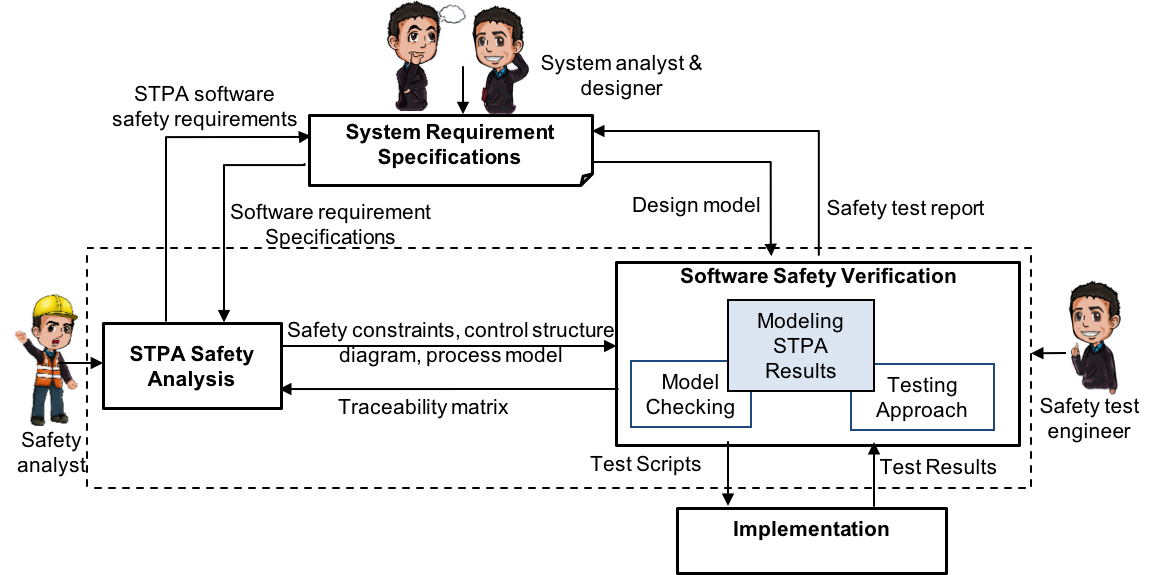}

\caption{STPA SwISs approach for software-intensive systems}

\label{fig0}

\end{figure*}

\subsection{STPA SwISs Approach for Software- Intensive Systems}

Developing safety-critical software requires a more systematic software and safety engineering process that enables the software and safety engineers to recognize the potential software risks. For this purpose, we proposed a comprehensive software safety engineering approach based on STPA for Software-Intensive Systems, called STPA SwISs \cite{Abdulkhaleq20152} (as shown in Fig. \ref{fig0}). STPA SwISs provides a concept of deriving the software safety requirements by STPA at the system level, modelling them as a safe behavioural model and automatically generating test cases from this model by using an existing model-based tool such as ModelJUnit \cite{Utting:2006:PMT:1200168}. The STPA SwISs approach is carried out in three major steps: 1) Deriving the software safety requirements at the system level, and generating the unsafe scenarios based on the extended approach to STPA by Thomas \cite{Thomas}, and expressing the corresponding safety requirements in formal specifications using LTL;  2) Modeling STPA results with a safe behavioural model. A safe behavioural model is a UML statechart notation that models the process model variables of a software controller in the STPA control structure diagram as states and the control actions as the state actions, and it is constrained by the STPA-generated software safety requirements (transitions); and 3) Generating the safety-based test cases by using an existing model-based tool.

Our preliminary algorithm \cite{Abdulkhaleq20152} for deriving test cases from STPA results relied  on using an existing model-based testing tool called ModelJUnit to drive the test cases. This algorithm is effective in deriving test cases but it has some limitations: 1) ModelJUnit requires that a behavioural  model be written as a Java class, which represents the finite state machine of the system; 2) The ModelJUnit tool has not been developed with the purpose of safety-based testing and deriving the test cases from the safety analysis results; 3) there is no way to verify and check the correctness of a test input model of ModelJUnit against the safety analysis results; and 4) The ModelJUnit tool does not provide a traceability matrix between the safety requirements and the generated test cases.

To support the safety engineering process based on STPA, we developed an extensible platform called XSTAMPP\footnote{\url{http://www.xstampp.de}} \cite{Abdulkhaleq2014a:2014} which is an open-source platform written in Java based on the Eclipse Plug-in-Development Environment (PDE) and Rich Client Platform (RCP). XSTAMPP supports performing the three main steps of STPA and provides an internal representation in XML for each STPA component to support possible future integration with other tools. STAMPP also supports the application of the \emph{STPA SwISs} approach in identifying the unsafe scenarios and automatically deriving the corresponding software safety requirements. It also supports the formal verification activities by automatically expressing the refined software safety requirements into formal specifications in LTL. As a new extension to XSTAMPP, we developed an Eclipse plugin called STPA verifier\footnote{http://www.xstampp.de/STPAVerifier.html} to automatically verify the refined software safety requirements which are expressed in LTL by using the model checkers (NuSMV and SPIN) in XSTAMPP.

In this paper, we discussed the algorithms to automate some activities of the \emph{STPA SwISs} approach which are implemented in XSTAMPP to help the software and safety engineers in identifying the unsafe scenarios, automatically deriving the corresponding software safety requirements, and automatically expressing the refined software safety requirements into formal specifications in LTL.

\subsection{Software Safety Testing}

Software testing is one of the most important phases during the software development process to detect inconsistencies between the software implementation and its requirements. A popular testing approach called Model-based Testing (MBT) \cite{Dalal:1999:MTP:302405.302640,Apfelbaum97modelbased} aims at automatically generating test cases using models extracted from software requirements. The model-based testing process involves creating a suitable model of the software's behaviour based on requirements or an existing specification to generate the test cases.

A big challenge in software testing is the design of test cases. To generate test cases, the tester needs first to understand the system specification and requirements. After that, the tester has to manually write test cases or automatically generate test cases from a model by using model-based testing tools. Automated generation of test cases involves that the system behaviour should be modelled in a suitable model. Over the years, there are many of the automated model-based test case generation approaches which have been developed by different techniques such as random generation algorithms \cite{1174916}, graph search algorithms \cite{1298776,Broy2005}, model-checking \cite{Offutt_generatingtest}, symbolic execution  \cite{Pretschner01classicalsearch} or theorem proving  \cite{Castanet2002}.

Software safety testing \cite{NASA2004,Lutz2000} is a crucial process in developing safety-critical systems to verify whether a software system meets its safety requirements. Safety-critical software should be tested extensively to ensure that the potential software-related hazards have been eliminated or controlled to a low level of risk.

\begin{figure}[t!]

\centering

\includegraphics[width=3.5in]{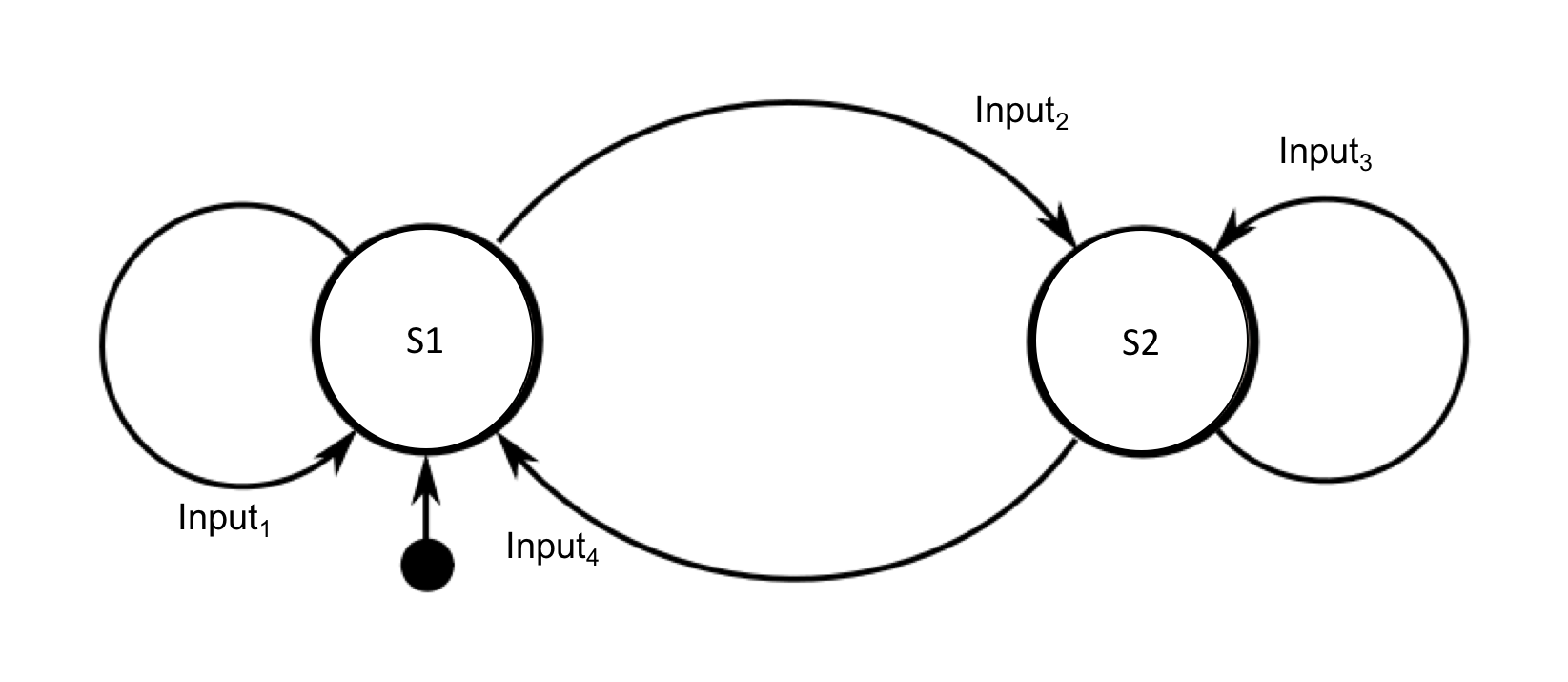}

\caption{The finite state machine model}

\label{fig:Figure2.9}

\end{figure}

A number of software behaviour models are in use today, several make good models for testing such as control flow charts \cite{Harel1987231}, finite state machines \cite{6771467,introduction}, SpecTRM-RL \cite{Leveson:2000:CFS:349360.351140}, and sequence event diagrams \cite{UML}. A software behaviour can be described as an input sequence, actions, guards and output logic, or the data flow through the software modules and routines. In the following, we describe popular software behaviour models which are used to model software behaviour and generate test cases from these models:

Finite state machines are commonly used in software behaviour modelling and testing to generate test cases \cite{Apfelbaum97modelbased}. The finite state model (shown in Fig. \ref{fig:Figure2.9}) includes a set of states, a set of input events and the transition between them.

\begin{definition} [Finite State Machine (FSM) ]

Let $f$ be a finite state machine which can be defined with a 5-tuple \cite{6771467}:

$f=(S, s_{1}, I, O, T_{s})$, where $S$ is a finite nonempty set of states with $s_{1}$ as the initial state, $I$ and $O$ are finite input and output alphabets, and $T_{s}$ is a behaviour relation which defines all possible transitions of the finite state machine model.  

\label{def: FSMformula}


\end{definition}

Another software behaviour model is an Extended Finite State Machine (EFSM) which is an extension of the classical (Mealy) Finite State Machine (FSM) model with input and output parameters, context variables, operation and guards defined over context variables and input parameters. EFSM \cite{1600197} is a common and very useful diagram to model the system behaviour and suitable for driving the test cases. EFSM contains nodes which represent the states of the system and the directed arcs which represent the transitions of the system from one state to another \cite{Utting:2006:PMT:1200168}.

\begin{definition} [Extended Finite State Machine (EFSM)]

Let $M$ be an extended finite state machine which can be defined by the 6-tuple \cite{1600197}:

$M=(Q, \sum_{1}, \sum_{2}, I, \vee, \wedge )$, where $Q$ is a finite set of states, $\sum_{1}$ is a finite set of events, $I \subset Q$ is the set of initial states, $\vee$ is the set of state variables, and $\wedge$ is a finite set of transitions.


\label{def: EFSMformula}

\end{definition}

Statecharts \cite{Harel1987231} were developed as a broad extension of the conventional formalism of finite state machines with notations of hierarchy, concurrency, and communication for describing the behaviour of complex or reactive software systems.

\begin{definition} [A Statechart] 	Let \emph{SC} be a statechart which can be defined with a 7-tuple  \cite{Harel1987231}:

$\mathit{SC}=(S, s_{1}, \lambda, \xi, \chi, \Omega, \Sigma )$, where $S$ is a finite set of superstates, $s_{1} \in S$ is as the start superstate which is a either a state or a state-chart, $\lambda$ is a transition function that maps the set of states $S$, $\xi$ is a superstate function that maps the set of superstates onto itself, $\chi$ is an event function that maps the set of transitions $T$, $S \times S$ to a set of events, $\Omega$ is a finite set of events, and $\Sigma$ is a default transition function that maps the set of states $S$ to their default sub-state if it exists, or itself. 

\label{def: statechart}

\end{definition}

In \cite{Harel1987231} Harel defined the statecharts language and the semantics of statecharts for complex systems. Simply, each Stateflow has a chart which is an independent state machine. Each chart has one or more states which are linked together by arcs labeled with transition information. The states can be also hierarchical states and contain a number of sub-states (children). Each state should have a type of state decomposition \emph{OR\_STATE} or \emph{AND\_STATE}. The \emph{OR\_STATE} decomposition allows only one sub-state (which has a default transition) to be active at a time when the parent (superstate) is active whereas the \emph{AND\_STATE} decomposition allows all sub-states to be active when the parent (superstate) is active.  

\begin{figure}[t!]

\centering

\includegraphics[width=4.0in]{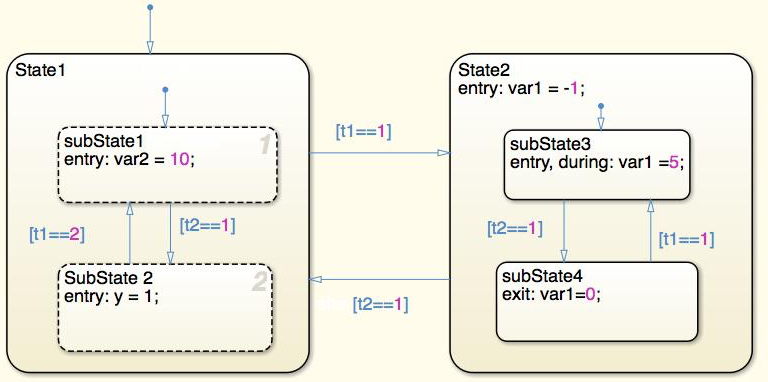}

\caption{The statechart model}

\label{fig:Figure2.6}

\end{figure} 

Based on Harel's statechart notations, the Simulink Stateflow model was developed by Mathworks\footnote{\url{http://www.mathworks.com}} to model event-driven (reactive) systems with enabling the representation of hierarchical state machine diagram, parallelism and history within statechart diagrams. The Simulink Stateflow is generally used to model the discrete controller in the model of a hybrid system where the continuous dynamics such as the behaviour of the plant and environment are specified using others capabilities of Simulink toolkit \cite{Hamon2007}. Recently, Matlab/Simulink has become a common model-based development tool for industrial software systems, which is widely used in various industries such as the aircraft, automotive, telecommunications, and transportation industries.

Figure \ref{fig:Figure2.6} shows the semantics of the Simulink's Stateflow model. In Simulink's Stateflow, each state can be labeled with the following elements:

\begin{tabular}{l} 
name $<$of a state$>$\\
entry: $<$entry actions are executed when a state is entered.$>$\\
during: $<$during actions are executed when while in a state.$>$\\
exit:  $<$exit actions are executed when a state is left.$>$\\
entry, during, exit: $<$combined actions in a state$>$
\end{tabular}

The Stateflow model in Fig. \ref{fig:Figure2.6} includes two superstates (state 1 and state 2) with state decomposition \emph{OR\_STATE}. The superstate \emph{State 1} has two sub-states with state decomposition \emph{AND\_STATE}. When the superstate \emph{State 1} is active, then the both sub-states will be active in the same time. While the superstate \emph{State 2} has two sub-states with state decomposition \emph{OR\_STATE}. That means when the superstate \emph{State 2} is active, then only the sub-state with the default transition \emph{subState 3 } will be active.

The semantics of the Stateflow are defined informally in Simulink. Therefore, the Stateflow model need to be translated into a formal model supported by the verification approaches. Two different semantics for the Stateflow were proposed: 1) denotational \cite{Hamon:2005:DSS:1086228.1086260} by Hamon and 2) formal operational \cite{hamon2007operational} by Haman and Rushby. Therefore, it is important here to note that our proposed algorithm to translate the Stateflow model into an SMV model works only with a subset of the operational semantics of the Stateflow model. Moreover, our algorithm does not consider the semantic of join transitions which are allowed in the Stateflow semantics.

\section{Related Work}

In the following, we will discuss the related work to our approach. 
\subsection {Risk-Based Software Testing}

There are several software safety test techniques in the literature that combine safety analysis principles with model-based testing. Most of them use the term ``Risk-based Testing", which combines risk analysis approaches such as FTA, FMECA or Markov chains with software testing approaches (e.g.\ model-based testing) to create a prioritization criterion for generating test cases.

Redmill \cite{Redmill:2004:ERT:1077269.1077272} explored the benefits of risk-based testing as the basis of test planning in the software testing process and how to understand the risks of the system to focus test efforts. He does not show how to generate the test cases from the risk analysis approach.

Zimmermann et al.~\cite {Zimmermann} proposed a refinement-based approach to the reliability analysis of safety-critical systems. They used statistical testing as a model-based testing technique and a Markov chain model to model the system under test. They also used FTA and FMECA as risk-based analysis techniques to identify the critical situations that represent high risk.

Kloos et al.~\cite{5954386} proposed a model-based testing approach which uses the information derived from FTA in combination with a system model to generate the risk-based test cases. They used FTA to select, generate and prioritize the test cases. They derived the test cases from the combination of fault trees and a basic system behaviour model called the ``base model".

Our approach uses a similar idea of combining a risk analysis approaches with model-based testing. The main difference is that we employ STPA for safety analysis which is based on system and control theory rather than reliability theory like FTA and FMEA. STPA copes with the analysis of complex, modern systems and tackles the dynamic behaviour of the system by treating safety as a control problem. Furthermore, STPA provides an abstract model of the system under analysis called the safety control structure diagram which views all main interacting components including the software components of the system. This allows us to directly construct the test model from the control structure diagram and constrain its transitions with the STPA-generated safety requirements.  

\subsection{Translating Simulink Models into Models Supported by the Verification Approaches}

A considerable amount of work has been done on translating Simulink models into models supported by formal verification approaches. In the following, we will discuss the most related work:

Banphawatthanarak et al.~\cite{banphawatthanarak1999symbolic} developed a tool called \emph{sf2smv} that generates input for the symbolic model checker SMV \cite{McMillan:1993:SMC:530225} from Stateflow models. In our work, we use the same concept for translating Simulink's Stateflow into the SMV model. As sf2smv is not available yet, however, it is difficult to compare it with our approach in detail. 

Meenakshi et al.~\cite{meenakshi2006tool} discussed the principles of translating Simulink models into an input language of a suitable model checker and providing reverse translation of traces violating requirements into Simulink notation for playback. They developed a translator from Simulink to the model checker NuSMV~\cite{NUSMV}. The translator takes a Simulink model as input and generates an equivalent NuSMV model. However, this translator is restricted only to discrete Simulink models and support only the basic blocks of Simulink (e.g. logical block or Selector block) that forms the finite state model of a system. Moreover, the translator does not support the translation of Simulink Stateflow into the input language of NuSMV.

Chen and Dong~\cite{DBLP:conf/icfem/ChenD06} proposed a systematic approach to translate Simulink diagrams to Timed Interval Calculus (TIC) \cite{Fidge:1998:SMR:648084.747169}, a notation extending Z to support real-time system specification and verification. The translated TIC specification covers the functional and timing aspects of the Simulink blocks. This work aims to guarantee the correctness of control systems. However, this work does not cover Stateflow diagrams. 

Chen et al.~\cite{DBLP:journals/sttt/Chen0LDZ12} proposed an approach to systematically translate Stateflow diagrams to into CSP\# \cite{sun2009integrating}. They developed a translator which is integrated inside the PAT model checker \cite{sun2009pat} to automate the process with support of different Stateflow features. This work aims to validate the functional correctness of Stateflow diagrams by detecting the bugs in the Stateflow model. The properties to be verified are declared in the CSP\# model with preprocessor such as \emph{\#define}. The translation process preservers the execution semantics of Stateflow and considers advanced Stateflow modelling features such as implicit events and history junctions.

Ferrante et al.~\cite{Ferrante2012} developed a modified tool called Parallel NuSMV (PNuSMV) based on NuSMV model checker \cite{NUSMV} that integrates the ManySAT parallel STA solver \cite{hamadi2008manysat}. This tool is part of the formal specification verification framework for the formal verification of Simulink/Stateflow models. The tool translates a subset of Simulink blocks (e.g.\ logical operators and arithmetic blocks) into the NuSMV meta model. The interesting properties are expressed as temporal logic to be verified with the PNuSMV tool. However, this work does not consider translating the Stateflow model into the NuSMV specifications.

In very recent work, Yang et al.~\cite{7582827} presented a tool for the translation of Stateflow models to timed automata to check the correctness of the Stateflow model. The translated model is used as an input to Uppaal timed automata \cite{Alur:1999:TA:647768.733787}. They used Uppaal to analyse the timing behaviour of the system and check both safety and liveness properties of timed automata.  We use a similar principle of transforming the Simulink's Stateflow model into an intermediate model with consideration of the state decomposition (AND\_STATE and OR\_STATE) and the attached actions (Entry, During and Exit). The main contribution of their work, however, is an approach to model check Simulink models which is not our main focus in this paper. Our contribution is to visualise the STPA process model, which is created during the safety analysis, with the Stateflow notation and check its correctness against the software safety requirements. In this way, we ensure that both models contain the same specifications (e.g.\ names of states, variables and control actions) before using it for generating test cases.

In conclusion, the existing work provide a great basis for our approach but are different in their focus. They concentrate on model checking Simulink models, not the integration with safety analysis or testing. To the best of our knowledge, there is no existing work on constructing the Stateflow diagram based on the information derived during a safety analysis for test case generation.

\subsection{Generating Test Cases  from Extended Finite State Machine}

Over the years, many approaches have been developed to generate test cases from statechart diagrams. The idea behind these approaches is to transform the statechart diagram into an Extended Finite State Machine (EFSM) and generate the test cases from this model. In the following, we will discuss the most related work: 

Ural \cite{URAL1987234} proposed a method to transform the extended finite state machines into a flow graph and generate test sequences. This method is based on the principles of using data flow analysis techniques in software reliability \cite{Fosdick:1976:DFA:356674.356676} to trace the flow graph and generate test cases.

Bourhfir et al.~\cite{Bourhfir1997} proposed a unified method for automatic executable test case and test sequences generation which combines both control and data flow testing techniques with control flow criteria (Unique Input Output) and the all-du paths coverages criterion. Their approach  generates only executable test cases for EFSM-specified systems by using symbolic evaluation techniques.

Kim et al.~\cite{Kim1999} proposed an approach to generate test cases from UML state diagrams based on the conventional control and data flow analysis. The authors have first transformed the state diagrams into EFSM with consideration of the hierarchical and concurrent structure of states (flattened and broadcast) of the UML state diagrams. Then, the EFSM are transformed into the flow graphs. They applied the conventional data flow analysis techniques to the resulting flow graph to generate the test cases. However, they focused only on identifying possible control and data flow and not the values of input variables. 

Hong et al.~\cite{STVR:STVR212} developed a method for the selection of test sequences from statecharts. The method is based on the STATEMATE semantics of statecharts by Harel \cite{Harel:1996:SSS:235321.235322}. The basic idea is to transform the statechart into an EFSM which contains all the possible runs of the statechart. The authors have considered the input variables in the EFSM which was generated from the state machine diagram. The resulting EFSM model will then be transformed into a flow graph to generate test sequences that cover all associations between definitions and uses of each variable that appear in the original state machine. The authors used the existing method of Ural \cite{URAL1987234} to transform the EFSM into a flow graph that models the flow of both the control and the data in the statechart.  

In conclusion, our approach uses a similar principle of generating test cases from the test model by using graph search algorithms (depth-first search, breadth-first search and both combined) which are presented in the aforementioned mentioned approaches. However, we choose different test coverages criterion to generate test cases such as all states coverage, all transition conditions coverage and the STPA software safety requirements coverage. However, our approach transforms each state in the safe test model as an executable Java script function that takes the state variables which are declared in the state actions (Entry, During, Exit) as parameters and execute their equations to update their values. The updated values of these variables will be used to check the transition condition and determine the next state. Moreover, our approach provides traceability between the software safety requirements and test model and traceability between the software safety requirements and the generated test cases.

\subsection{Generating Test Cases from Simulink Models}

Few research has concentrated on the subject of the automatic generation of test cases from Simulink Stateflow:
Zhan and Clark \cite{Zhan:2008:SFA:1326359.1326426} developed an approach for automatic testing of Matlab/Simulink models. Their approach is a search-based test data generation and selection approach. It uses the first search-based approach to generate test data.

P\^{a}s\^{a}ranu et al.~\cite{5226844} proposed a framework for model-based analysis and test case generation for flight software of a NASA flight mission based on the Simulink Stateflow and UML representations. They used Java path finder\footnote{\url{http://javapathfinder.sourceforge.net}} and symbolic path finder\footnote{\url{http://babelfish.arc.nasa.gov/trac/jpf/wiki/projects/jpf-symbc}} to generate test cases from both UML and Simulink/Stateflow models. The proposed framework is based on the concept of using model checking to generate test cases. The framework takes the models which are created by using different modelling environment and enables their analysis with model checking and test case generation approaches. 

Windisch \cite{5071030} proposed an automation approach for search-based testing of continuous functional models like Simulink Stateflow models. The method demonstrates how search-based testing techniques can be applied to a continuous functional model such as Simulink/Stateflow to generate test cases.

Li and Kumar \cite{6386487} proposed an automatic method for test data generation for Simulink Stateflow based on its translation to input/output extended finite automata model. The method involves that the Simulink Stateflow shall translate to an input/output extended finite automate model. Each path in Input/output extended finite automate model represents a computation sequence of the Simulink Stateflow diagram. They implemented the proposed method by using two model checking techniques and constraint solving. The NuSMV model checker is used to map the input/output of the extended finite automata to a finite abstracted transition system modeled in SMV and generate test cases by checking each path in the I/O of the extended finite automata against the resultant model.  

We differentiate our work here from the aforementioned approaches by automatically generating the safe test model from the Stateflow model which is constructed from the safety analysis specification. Our approach also shows how to validate the correctness of the safe test model against the software safety requirements by using the NuSMV model checker.

 \begin{figure*}[t]
 	
 	\centering
 	
 	\includegraphics[ ]{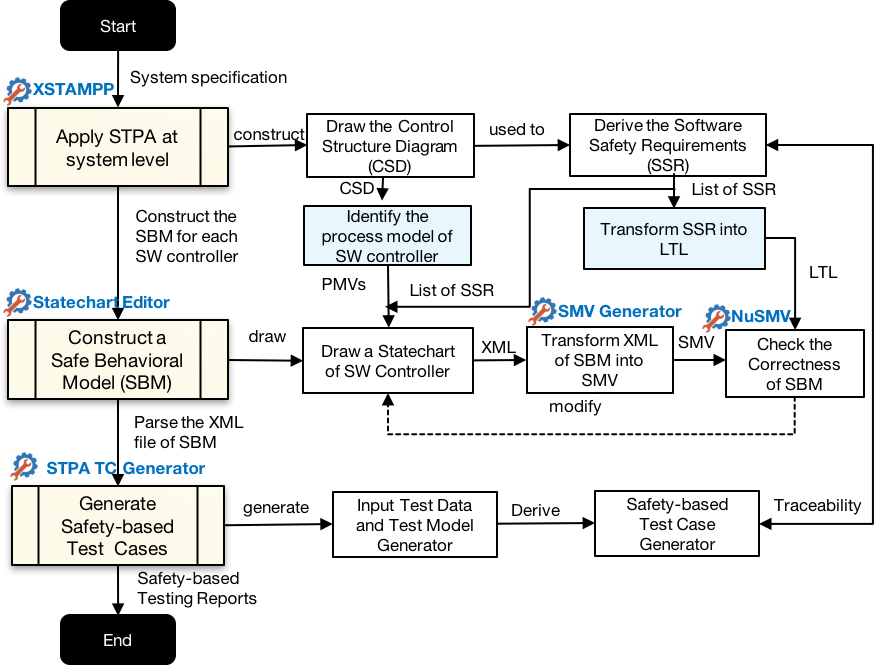}
 	
 	\caption{An overview of the proposed approach}
 	\label{proposedapproach}
 \end{figure*}

\section{The Proposed Approach}

In this section, we propose an automatic safety-based test case generation approach for deriving test cases directly from the STPA safety analysis results. The proposed approach follows the main steps of the STPA SwISs approach and provides a high degree of automation of each step. 

We introduce a running example that we will use to illustrate our approach. It is the software controller of a train door system. Let us assume the software controller of the train door control system was designed to open and close a door of a train and monitor the status of the door. The controller works by receiving information about the position and the status of the door by a sensor. The door controller also receives input from external sensors about the position of the train and whether an emergency is happening. Then, the controller issues the door open or close commands. The actuator will receive these commands and apply mechanical force on the physical door. 

Figure \ref{proposedapproach} shows the main steps of the approach which includes four steps: 1) deriving the software safety requirements of a software controller by following the STPA SwISs approach \cite{Abdulkhaleq20152} and automatically expressing them in LTL, 2) constructing the safe behavioural model of the software controller with the statechart notations in Simulink, 3) transforming the safe behavioural model into an input model of the NuSMV model checker and checking the correctness of the generated model against the STPA and safety requirements expressed in LTL; and 4) automatically generating a safety-based test model and deriving  the safety-based test cases from this model.

In the following sections, we describe the four major activities of the proposed approach in more detail.
 
\subsection {Deriving Software Safety Requirements}

This step starts by applying STPA to the system specification to identify STPA software safety requirements and the potentially unsafe scenarios which the software can contribute to. The algorithm starts by establishing the fundamentals of analysis by determining the system-level accidents (\emph{ACC}) and the associated system-level hazards (\emph{HA}) which the software can lead to or contribute in. Next, the algorithm demands that the safety control structure diagram of the system shall be constructed from the system specifications. The software here is the controller in the control structure diagram.

\begin{figure}[t]
 	\centering
 	\includegraphics[width=4.5in]{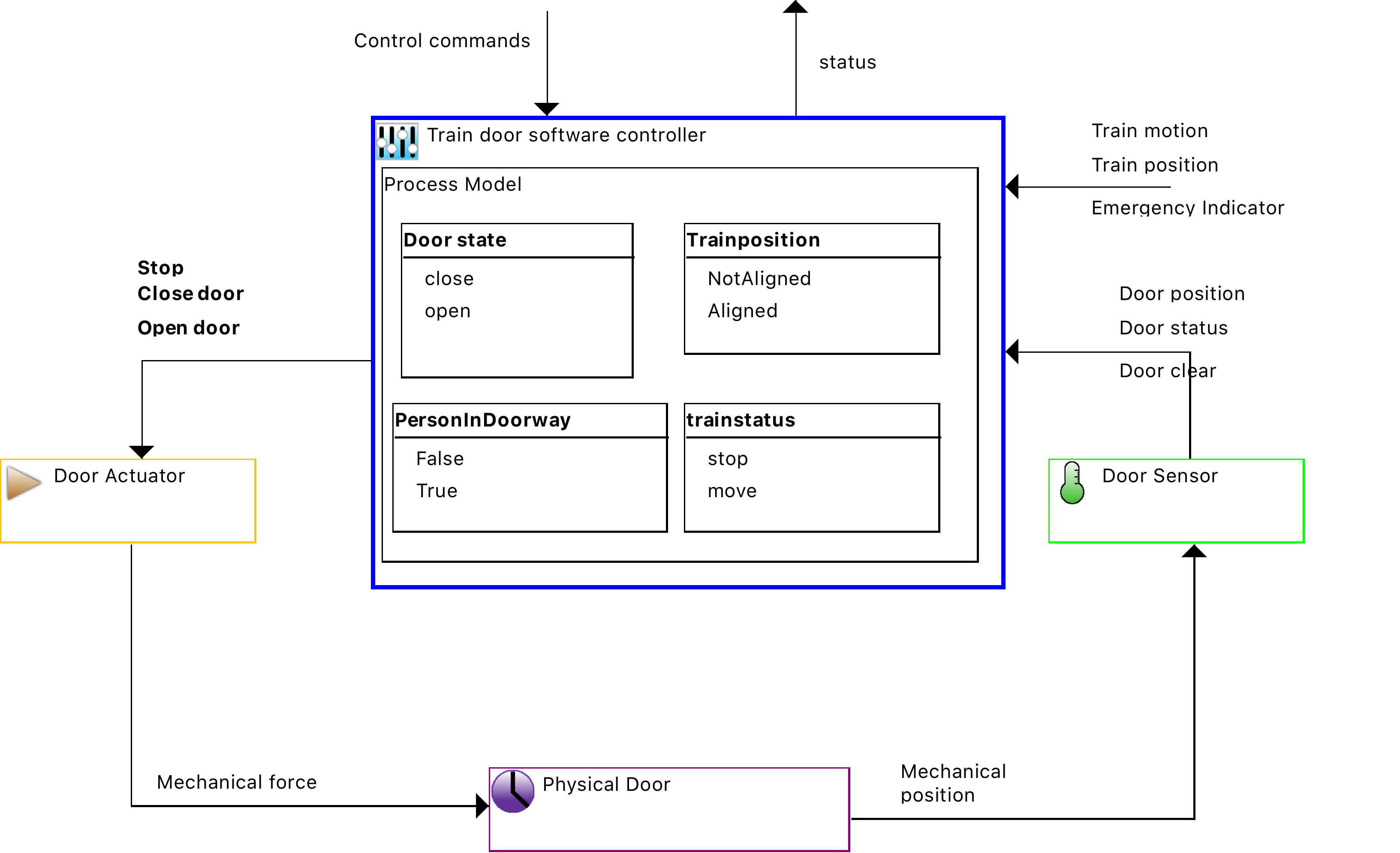}
 	\caption{The control structure diagram of train door system with the safety-critical process model variables}
 	\label{figtrainACControlstructure}
 \end{figure}

For our running example of the train door, we can reuse the STPA analysis in \cite{Thomas2011}. For example, they came up with the accident \emph{ACC.1}: \emph{A person is injured while the train closed the door}. A system-level hazard that can lead to this accident is, for example, \emph{HA.1}: \emph{Door closed the door while a person is in the doorway}.
The control structure diagram of the train door system is shown in Fig.~\ref{figtrainACControlstructure}). The control structure diagram includes: 1) software door controller; 2) door actuator; 3) physical door; 4) and door sensor.

For each software controller component in the control diagram, its software safety requirements can be derived by performing the following steps:

\begin{enumerate}
	
	\item STPA Step 1: Identify unsafe control actions.  In this step, the safety analyst will identify the potentially unsafe software control actions for each software component that can lead to one or more of the defined system hazard \emph{HA}, as follows:
	
	\begin{enumerate}

		\item Identify all safety-critical Control Actions (\emph{CA}s) that can lead to one or more of the associated hazards (\emph{HA}). For example, the software controller of the train door system issues three control actions: \emph{open door, close door, stop opening or closing door.}

		\item	Evaluate each CA with four general types of hazardous behaviours to identify the Unsafe Control Actions (\emph{UCA}s): (a) a control action required for safety is not provided, (b) an unsafe action is provided, (c) a potentially safe control action is provided too early, too late or out of sequence and (d) a safe control action is stopped too soon or continued too long. For example, the control action \emph{CA.1:} \emph{close door} can be unsafe and lead to the hazard \emph{H1}. It is then \emph{UCA1.1}: \emph{The software controller closes the door while a person is in the doorway}.

		\item Translate the identified \emph{UCA}s manually into informal textual Software Safety Requirements (\emph{SSR}). For example, the corresponding software safety requirement of \emph{UCA1.1} is \emph{SSR1.1:} \emph{The door software controller must not close the door while a person or object is in the doorway.}

		\item Identify the process model and its variables and include them in the software controller in the control structure diagram to understand how each \emph{UCA} could occur. The process model describes the states of the software controller (only critical states which are relevant to the safety of the control actions) and their variables describe the software communication, input and output. Figure \ref{figtrainACControlstructure} already shows the process model of the software controller. The process model has four process model variables \emph{PMV}: 1 state variable $S$ ( $S_{1}=$ door state) and 3 variables \emph{P}($P_{1}=$personInDoorway, $P_{2}$=Train status and $P_{3}$=Trainposition)

		\item Automatically generate the critical set of combinations of the process model variables for each control action ($\mathit{CA}$). Each combination should be evaluated within two contexts ($C_{1}$= \textbf{Providing $\mathit{CA}$} or $C_{2}$  = \textbf{Not Providing $\mathit{CA}$}) to determine whether the control action is hazardous in that context or not. A control action $\mathit{CA}$ could be considered hazardous in context $C$ if only a combination of process variables related to $CA$ leads to a system-level hazard $H \in \mathit{HA}$. The context $C_{1}$ = \textbf{Providing $\mathit{CA}$} has three types of sub-contexts: \emph{context incorrectness}, in which the unsafe control action commanded incorrectly and caused a hazard (any time), \emph{context real-time execution}, in which the unsafe control action commanded in a wrong timing (too early or too late) or sequence, and \emph{context execution mechanism}, in which the unsafe control action commanded in a wrong mechanism of execution (applied too long or stopped too soon).  For example, the process model variables that have an effect on the safety of providing or not providing the control action \emph {CA1:} \emph{close door} are \emph{door state, train position, person in door and train status}. An example of the critical set of combinations which can be generated based on the process model variable values for the \emph{close door} control action in the context of providing $C_{1}$ is $Cs_{1}:$ \emph{door state = open,  train position = aligned, a person in doorway = yes and train status = stop}
		
		\end {enumerate}

		\item STPA Step 2: Identify the unsafe software scenarios for each unsafe control action. Based on the results of STPA Step 1, the safety analyst will identify the unsafe software scenarios for each unsafe control action \emph{UCA} as follows:

		\begin{enumerate}

			\item Identify the potentially unsafe critical combination of unsafe software control actions and evaluate it to identify the potential unsafe scenarios of the software controller that cause accidents. For example, an unsafe scenario which can be derived from the critical combination $CS_{1}$ is \emph{SC1.1:} \emph{The door software controller provided the control action \emph{close door} while the train is stopped and train position is aligned, door state is open and a person is in the doorway}.

			\item Refine software safety constraints based on the unsafe scenarios of the software controller. For example, the software safety requirement which can be formulated from the unsafe scenario \emph{SC1.1} is \emph{SSR1.1:} \emph{The door software controller must not provide the close door command while the train is stopped, the train position is aligned with the platform, the door state is open and a person is in the doorway}

		\end{enumerate}

	\end{enumerate}

\begin{definition} [Refined Unsafe Control Action] The refined unsafe control action ($\mathit{RUCA}$) is a four-tuple ($\mathit{CA}, Cs, C, \mathit{TC}$), where $\mathit{CA}$ is a control action which causes a hazard $H\in \mathit{HA}$, $Cs$ = $\bigcup ({\cal P}_{1}=v_{1}, \ldots {\cal P}_{n}=v_{n})$ which is a critical set of combinations of the relevant process model variables $\mathit{PMV}$ of $\mathit{CA}$, $C$ is a context where  providing or not providing the control action $CA$ is hazardous, and $TC$ is the type of context \textbf{providing} of control action $\mathit{CA}$ (\textbf{any time}, \textbf{too early} or \textbf{too late}). 

\end{definition}

To automatically translate each critical combination of process model variables for each control action $CA$ into the unsafe software scenarios, we set the following rules:  \\

  \textbf{Rule 1: }Each refined unsafe control action ($\mathit{RUCA}$) in the context of \textbf{Providing} ($C_{1}$) of a control action $CA_{i}$ can be expressed as: \\ 

$\mathit{RUCA}_{i}$ = $<$CA$>$ \textbf{provided} $<$TC$>$ \textbf{is hazardous when} 	$<$Cs= $\bigcup ({\cal P}_{1}=v_{1}, \ldots {\cal P}_{n}=v_{n})$$>$ occurred. \\

\textbf{Rule 2: }Each refined unsafe control action (RUCA) in the context of \textbf{Not Providing} ($C_{2}$) of a control action $\mathit{CA}_{i}$ can be expressed as:\hfill \\

$\mathit{RUCA}_{i}$ = $<$CA$>$ \textbf{Not provided} \textbf{is hazardous when}  $<$Cs= $\bigcup ({\cal P}_{1}=v_{1}, \ldots {\cal P}_{n}=v_{n})$$>$ occurred. \\

By using the rules 1 and 2, we refine the unsafe control actions which are identified based on the combination set of process model variables. The software safety requirements are generated automatically from the refined unsafe control actions. Based on definition 3, we identify the following rules which are used to automatically generate the Refined Software Safety Requirements ($\mathit{RSSR}$): \\

\textbf{Rule 3: }Each $\mathit{RUCA}_{i}$ in the context \textbf{Providing} ($C_{1}$) of control action $\mathit{CA}_{i}$ can be transformed automatically into a new software safety requirement as follows:  \\

$\mathit{RSSR}_{i}$ = $<$CA$>$ \textbf{must Not be Provided} $<$TC$>$  \textbf{when} $<$Cs= $\bigcup ({\cal P}_{1}=v_{1}, \ldots {\cal P}_{n}=v_{n})$$>$ occurred. \\

\textbf{Rule 4: }Each $\mathit{RUCA}_{i}$ in the context \textbf{Not Providing} ($C_{2}$) of control action $\mathit{CA}_{i}$ can be transformed automatically into a new software safety requirement as follows:   \\

$\mathit{RSSR}_{i}$ = $<$CA$>$ \textbf{must be Provided}  \textbf{when} $<$Cs= $\bigcup ({\cal P}_{1}=v_{1}, \ldots {\cal P}_{n}=v_{n})$$>$ occurred.

\subsection {Automatically Formalizing Safety Requirements in Linear Temporal Logic (LTL)} 
In \cite{Abdulkhaleq2015}, we described an algorithm to formalize the safety requirements in LTL. Here, we extend it to include software safety requirements that include timing. By using rules 3 and 4, each refined software safety requirement \mbox{$\mathit{RSSR}_{i}$}, which is identified from the refined unsafe control action $\mathit{RUCA}_{i}$, can be transformed automatically into a formal specification in LTL.

Rule 3 defines three types of software safety requirements, which means that the control action $\mathit{CA}_{i}$ must not be provided in the type of context $TC$ = \textbf{any time , too early or too late} when the critical  combination $\mathit{Cs}_{i}$ of the relevant process model variable values occurred. Each type of software safety can be transformed automatically into formal specification by the following rules:  \\

\textbf{Rule 3.1: }Each $\mathit{RSSR}_{i}$ derived from the context of providing control action $\mathit{CA}_{i}$ \textbf{any time} (without delay) can be  automatically transformed into LTL as: \\

$\mathit{LTL}_{i}$ = G ($Cs_{i}$ $\rightarrow$ !  ($controlAction$ == $\mathit{CA}_{i}$)),	where   $\mathit{Cs}_{i}$= $\bigcup ({\cal P}_{1}=v_{1}\wedge \ldots {\cal P}_{n}=v_{n} )$. 		\\

Rule  3.1 means that it always ($G$) the software controller should not ($!$) provide a control action $\mathit{CA}_{i}$ when the values of the critical combination $\mathit{Cs}_{i}$ have been occurred. \\

\textbf{Rule 3.2: }Each $\mathit{RSSR}_{i}$ derived from the context of providing control action $\mathit{CA}_{i}$ \textbf{too early} can be automatically transformed into LTL as: \\

 $\mathit{LTL}_{i}$ =  G ((($controlAction$ == $\mathit{CA}_{i}$) $\rightarrow$   $\mathit{Cs}_{i}$ ) \&   ! ( ($controlAction$ == $\mathit{CA}_{i}$)  $U$ $\mathit{Cs}_{i}$)). \\

Rule 3.2 means that a software controller should always ($G$) not provide control action $\mathit{CA}_{i}$ before the occurrence of critical combinations set $\mathit{Cs}_{i}$ still not become true in the execution path and that it well provides the $\mathit{CA}_{i}$ when the combination of $\mathit{Cs}_{i}$ holds. \\

\textbf{Rule 3.3: } Each $\mathit{RSSR}_{i}$ derived from the context of providing control action $\mathit{CA}_{i}$ \textbf{too late} can be automatically transformed into LTL as: \\

 $\mathit{LTL}_{i}$=  G$((\mathit{Cs}_{i} \rightarrow  (\mathit{controlAction} == \mathit{CA}_{i})) ~\& ~  ! (\mathit{Cs}_{i}   U (\mathit{controlAction} == \mathit{CA}_{i})) ).$ \\

Rule 3.3 means that the software controller should always ($G$) not provide a control action $\mathit{CA}_{i}$ too late while the occurrences of the critical set of combinations has become previously true in the execution path. \\

Rule 4.1 defines one type of the software safety requirements which is the context of not providing a control action $\mathit{CA}_{i}$ when it is required. This type can be expressed into LTL by the following rule: \\

\textbf{Rule 4.1: }Each $\mathit{RSSR}_{i}$ derived from the context of \textbf{Not providing} of control action $\mathit{CA}_{i}$ can be automatically transformed into LTL as:  \\

$\mathit{LTL}_{i}$ = G $(\mathit{Cs}_{i} \rightarrow X (\mathit{controlAction}==\mathit{CA}_{i}))$,	where  $\mathit{Cs}_{i}$= $\bigcup ({\cal 	P}_{1}=v_{1}\wedge \ldots {\cal P}_{n}=v_{n} )$.

This rule means that the occurrence of a critical set of combination values always implies that the software controller must provide the control action $\mathit{CA}_{i}$ at the next time step ($X$) without any delay.

For example, the corresponding LTL of the software safety constraint $SSR1.1$ of the train door software controller can be specified as follows: 
 
  $LTL_{1.1}$=   G (((trainstatus== stop) \&\& (doorstate $==$ close) \&\& (trainposition$==$ Aligned)  \&\& (PersonIndoorway$==$ TRUE))  $\rightarrow$ !  (controlAction$==$ closedoor)).

	\begin{figure}[t]
		
		\centering
		
		\includegraphics[width=4in]{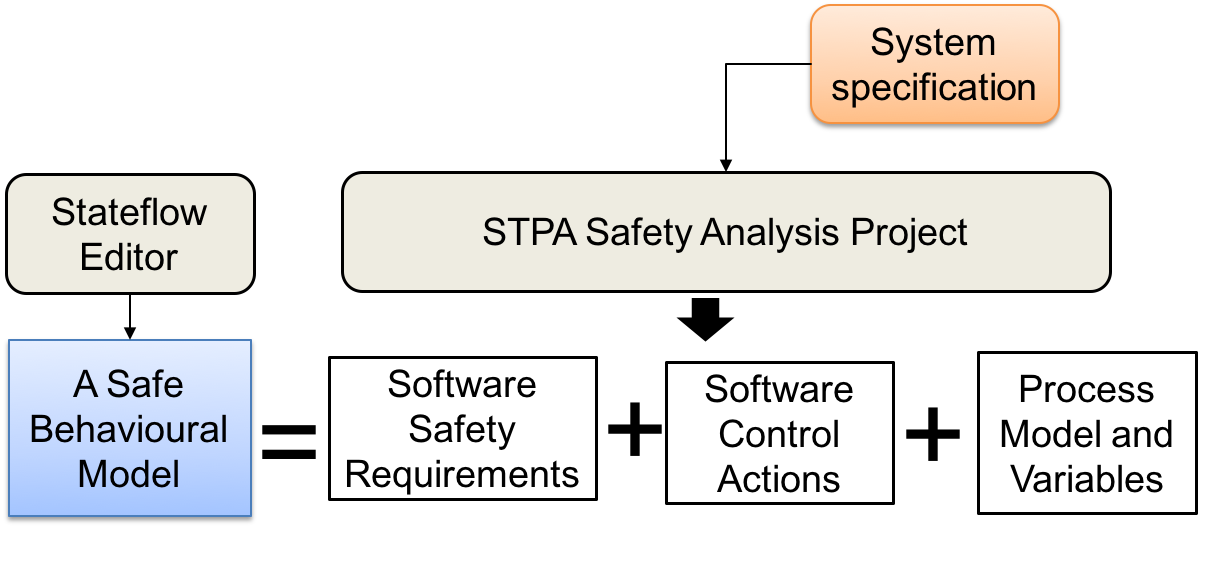}
		
		\caption{A safe software behavioural model}
		\label{figsafe}
	\end{figure}

\subsection {Constructing a Safe Software Behavioural Model} 

To generate the safety-based test cases, the information derived from the STPA safety analysis must be integrated into a suitable model which should visualise the process model variables of each software controller and their relations in a control structure diagram. For this purpose, we select the Stateflow \cite{Math2016} diagram notations to visualise the automation model of each software controller. The Stateflow diagram is a visual notation for describing dynamic behaviour, including the hierarchy, concurrency and communication information. The idea here is to build a model from STPA results with a modelling editor (e.g. Simulink) that supports the export of the statechart notations as XML specifications.   

\begin{figure}[t]
	
	\centering
	
	\includegraphics[width=5.5in]{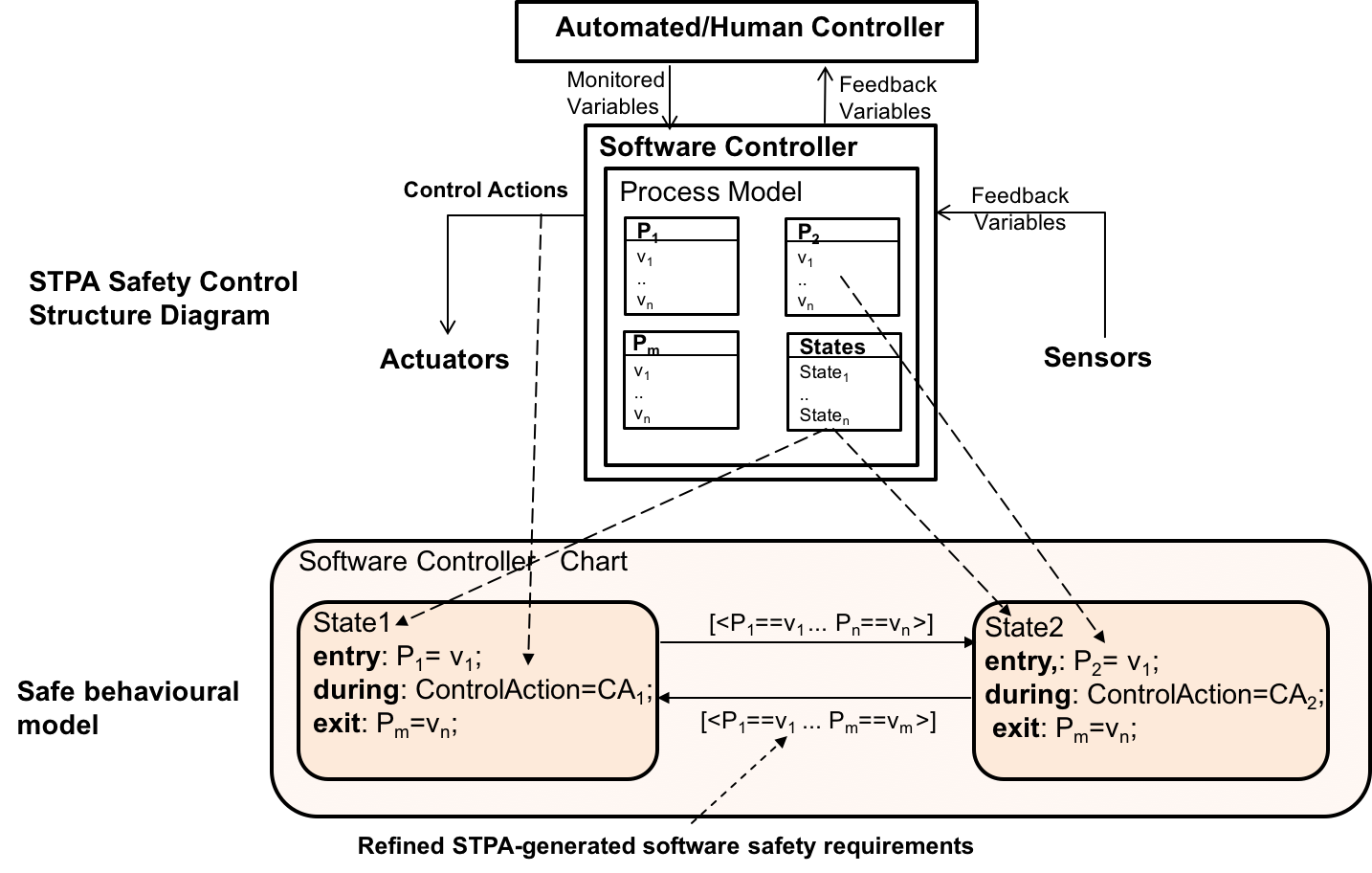}
	
	\caption{Mapping the process model variables and control actions into the safe software behavioural model}
	\label{fig9}
	
\end{figure}

\begin{definition} [Safe Behavioural Model (SBM)] Let $\emph{SBM}$ be a Safe Behavioural Model (shown in Fig. \ref{figsafe}) which can be expressed by a three-tuple ($\mathit{PMV}, T, CA$), where $\emph{PMV}$ is a set of the safety-critical process model variables: critical variables $P$ and $S$ states: $P \subset \mathit{PMV}$ and states $S \subset \emph{PMV}$, $T$ is the set of transition conditions which are extracted from the STPA refined software safety requirements $\emph{RSSR}$ that are refined based on $\emph{PMV}$, and $\emph{CA}$ is the set of the critical software control actions.

\end{definition}

Each transition $T_{i}$ of the safe behavioural model is expressed with the syntax \emph{$T_{i}$ = IE [SSR] / TA}, where \emph {IE} is the input event that causes the transition $T_{i}$, \emph{SSR} is a safety requirement which is a Boolean condition that constrains the transformation from the current state to the next state, and \emph{TA} is an action that will be executed when the Boolean expression is valid. Each state in the Stateflow model has three optional types of actions: \emph{Entry, During and Exit} actions. Entry actions execute when the state is entered, \emph{During} actions execute when the state is active, an event occurs and no valid transition to another state is available, and \emph{Exit} actions execute when the state is active and a transition out of the state occurs \cite{Math2016}. These actions are used to determine how to change the current state of the software controller to the next state.

 The syntax of the Stateflow in Simulink allows to combine these three actions that execute the same tasks in a state. To change values of the process model variables $P = $$\bigcup ({\cal 	P}_{1}=v_{1}\wedge \ldots {\cal P}_{n}=v_{n} )$ in a state, we used these actions of each state in the safe behavioural model to determine how each value of the process model variable ($P_{i}$) can be changed when the software controller enters or exists or this state. For example, the process model variable $P_{i}$ in the process model of the software controller can be written in a state as an Entry, During or Exit action or combined state actions as follows:  \emph{$entry, during, exit: P_{i}$ = $<$new value of $P_{i}$ $>$}. 

As the transition condition is derived from the refined STPA software safety constraints, the new value of each process model variable will be used to check the transition condition of the current state to determine what is the next state. We also used these state actions to determine which control action of the software controller can be dispatched on entering, during or exiting the current state. Figure \ref{fig9} shows how to map the internal process model variables of the software controller and its control actions into the safe behavioural model. We identify the rules of constructing a safe software behavioural model from the STPA results as follows:

 \begin{figure*}[t]
 	\centering
 	\includegraphics[width=3.5in]{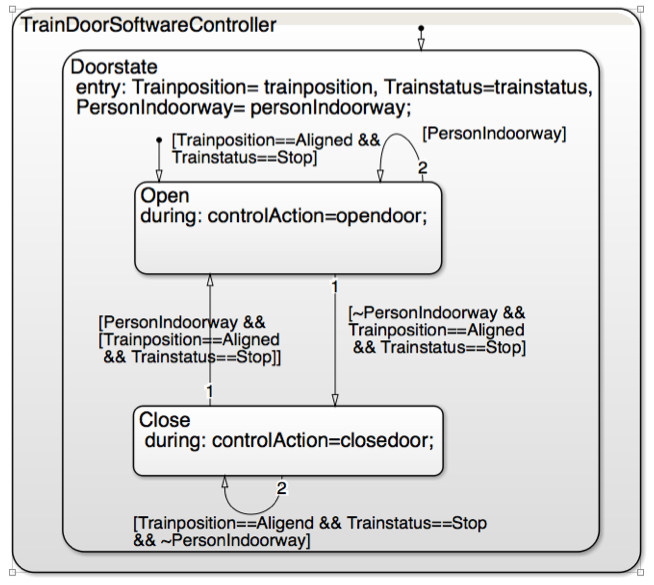}
 	\caption{The safe behavioural model of the train door software controller}
 	\label{safebehavoiral2}
 \end{figure*}

\begin{itemize}

\item The process model of each software controller in the STPA control structure diagram will be visualised by a chart in the Stateflow model. For example, the Stateflow model of the door software controller (as shown in 
Fig. \ref{safebehavoiral2}) has one chart which contains one superstate \emph{door state}.

\item The safe behavioural model should contain all internal states $S$ and the safety-critical process variables $P$ of the software controller in the STPA control structure diagram. As shown in figure \ref{safebehavoiral2}, the door software controller contains one internal state 
\emph{doorstate} which has two substates \emph{close} and \emph{open}  .

\item All process model variables of the software controller in the STPA control structure diagram should be declared in the safe behavioural model. As an example, we declared the three process model variables of the train door software controller as follows: Trainposition (Enum ), Trainstatus (Enum) and the personIndoorway (Boolean)

\item The safe behavioural model should constrain the transitions using the STPA software safety requirements (constraints) which are identified based on the rules 3 and 4. For example, the train door software controller must provide the control action \emph{open door} when the train is stopped and aligned with platform. We used this information to constrain the default transition of the state \emph{open}.

\item Define an enumeration data type variable named \emph{controlAction} in the safe behavioural model which takes all control actions of the software controller in the STPA control structure diagram as its value. For example, we defined the control actions of the train door software controller \emph{open door, close door} and \emph{stop} as enumeration data type in the safe behavioural model.

\item The \emph{controlAction} variable will be used as an entry/during/exit action of internal states of the safe behavioural model to show which control action will be issued when the software controller enters or exits a state.

\end{itemize}

\subsection {Automatically Transforming a Safe Software Behavioural Model into an SMV Model} 

To check the correctness of the safe behavioural model and ensure that the safe behavioural model of the software controller satisfies all STPA software safety requirements, the safe behavioural model must be verified against the generated LTL formulae. For this purpose, we developed an algorithm that automatically transforms the SBM model created in the Simulink editor into an input language of a model checker such as SMV (Symbolic Model Verifier), automatically parses the LTL formulae from the STPA data model and includes them into an SMV model. To verify the SMV model against the STPA software safety requirements, we use the NuSMV model checker. In case that the SMV model does not satisfy a given LTL of a software safety requirement, the NuSMV model checker will return a counterexample. A counterexample contains information that shows why the given LTL formula of a software safety requirement is not satisfied. Based on the counterexample's information, the safe behavioural should be modified. As the LTL formula contains information about the state of software controller $s_{i}$ and the control action $CA$, therefore, the modification of the safe behavioural model involves changes to the transition conditions or the initial values of the variables of the state $s_{i}$ in which the model violated the given LTL formula. This step continues until the safe behavioural model satisfies all STPA-generated software safety requirements. 
 
The algorithm of generating the SMV model is divided into three sub-algorithms: 1) \emph{generate STPA data model} which parses XML specifications of the STPA project created in XSTAMPP (shown in algorithm 1); 2) \emph{generate Stateflow (safe behavioural model) data model } which parses XML specifications of a Stateflow model and generate a tree of Stateflow states ($TSf$) in which a node represents one Stateflow state (shown in algorithm 2); and 3) \emph{generate SMV model } which transforms the STPA data model and Stateflow data model into SMV specifications  (shown in algorithm 3).  

\begin{algorithm}[t]

\caption{Generate STPA Data Model} 

\textbf{Input:} $F$ : A STPA project file 

\textbf{Data:} \\~~~~~~~~~ $CAs$= a list of control actions, \\~~~~~~~~~ $PMVs$= a list of process model variables, \\~~~~~~~~~ $SSRs=$ a list of software safety requirements, and \\~~~~~~~~~  $LTLs=$ a list of generated LTL formulae of $SSR$. \\~~~~~~~~~ $DataModel$= a data model which stores all information of STPA project $F$.

\textbf{Output:} $DataModel_{SW}$= a list of the data model of the software controller $CO$ $\in$ $F$. 

\textbf{Description:}

\begin{algorithmic}[1]

\State \textbf{URL schemaFile ($``/hazschema.xsd"$)} 
\State \textbf{XSModel} LoadXMLSchema ($schemaFile$)  
\State \textbf{DataModel} ParseXMLSchema ($F$) 

\For {\textbf{each} $SW_{i}$ Controller in $DataModel$} 

\State ~~\textbf{Create} a new data model $DM_{SW}$ for $SW_{i}$ Controller. 

\State ~~\textbf{Fetch:} $ CAs$ $\leftarrow$ $DataModel.fetchControlActions()$
\State ~~\textbf{Fetch:} $ PMV$ $\leftarrow$ $DataModel.fetchProcessModelVariables()$
\State ~~\textbf{Fetch:} $ SSRs$ $\leftarrow$ $DataModel.fetchSoftwareSafetyRequirements()$
\State ~~\textbf{Fetch:} $ LTLs$ $\leftarrow$ $DataModel.fetchLTLs()$

\State ~~\textbf{Add} $\mathit{DM}_{SW}.CAs$ $\leftarrow$ $CAs$
 \State ~~\textbf{Add} $\mathit{DM}_{SW}.PMVs$ $\leftarrow$ $PMVs$
\State ~~\textbf{Add} $\mathit{DM}_{SW}.SSRs$ $\leftarrow$ $SSRs$
  \State ~~\textbf{Add} $\mathit{DM}_{SW}.LTLs$ $\leftarrow$ $LTLs$
\State ~~\textbf{Add} $DataModel_{SW}[i]$ $\leftarrow$  $\mathit{DM}_{SW}$.

\EndFor

\State \textbf{Return} $DataModel_{SW}$

\end{algorithmic}

\end{algorithm}

\subsubsection{\textbf{Parsing the STPA project created by XSTAMPP}} Algorithm 1 shows the process of parsing the STPA project created by XSTAMPP. The algorithm process accepts the STPA project file $F$ as input. Then, it parses the XML specification of the STPA project into the corresponding data model $DataModel$ which represent all data in an STPA project (see lines $1-3$). For each software controller in the control structure diagram, a data model $DM_{SW}$ will be created to store the information about the software controller such as its critical control actions, process model and its variables, software safety requirements and the generated LTL formulae (see lines $4-5$). The algorithm will fetch the information of each software controller and store them in the corresponding lists (see lines $6-9$) and add these lists into the data model of the software controller (see lines $11-14$). The output of this algorithm is a list of the data model of the software controllers in the safety control structure diagram (see line $16$). 

\subsubsection{\textbf{Parsing the Stateflow model created by Simulink/Matlab}}
Algorithm 2 shows how to parse the XML specifications of the Stateflow model stored in a Simulink/Matlab file. The input of this algorithm is an XML file of the Simulink Stateflow file ($Sf$) which contains XML specifications of the Stateflow model. 

\begin{algorithm}[t]

\caption{Generate a Tree of Stateflow Data} 

\textbf{Input:} $Sf$ : A Simulink Stateflow file  

\textbf{Data:} $\mathit{DM}_{Sf}$ = A data model to store all data of Stateflow in $Sf$, \\~~~~~~~~~  $S$= A list of states of Stateflow $Sf$. 

\textbf{Output:} $T_{Sf}$= a tree which represents all information of Stateflow states $\in$ $sf$. 

\textbf{Description:}

\begin{algorithmic}[1]
\State \textbf{URL schemaFile ($``/Stateflowschema.xsd"$)} 
\State \textbf{XSModel} LoadXMLSchema ($schemaFile$)  
\State \textbf{$\mathit{DM}_{Sf}$} ParseXMLSchema ($Sf$) 

\State\textbf{Extract all states at level 0:}  $S$   $\leftarrow$   $\mathit{DM}_{Sf}$.Stateflow.getStates()

\State\textbf{Create a state root node }$\leftarrow$ $root$  

\State \textbf{Set ParentID}  $root$  $\leftarrow$ ParentID $\notin$ $\mathit{DM}_{Sf}$.$States.IDs$

\State \textbf{Name}  $root$  $\leftarrow$ 'root'

\For {\textbf{each State}  $s$  \textbf{in} $S$} 

\State~~~\textbf{Create a state child node }$\leftarrow$$node$  

\State~~~\textbf{Set ParentID}  $node.parentID$  $\leftarrow$  $root.ID$.

\State~~~\textbf{Set Data}  $node.name$  $\leftarrow$  $s.name$

\State~~$node.Id$  $\leftarrow$  $s.SSID$

\State~~$node.setDecomposition$ $\leftarrow$  $s.getDecomposition()$ 
\State ~~$node.setStatesActions$ $\leftarrow$ $s.getStatesActions() $//Entry, During and Exit Actions

\If {$s.hasChildren()$==true}

\State ~~~~\textbf{ $node.isHasChildren( true )$}

\State~~~~\textbf{traverseChildren ($node$ , $s$ )}

\EndIf

\State~~~~\textbf{Add $root.addChild$ ( $node$ )}

\EndFor 

// 	Extract all transitions between the states.	

\State~~~~\textbf{$T_{Sf}$.setTransitions($\mathit{DM}_{Sf}.getTransitions()$)}	 

// 	Extract all variables of Stateflow.

\State~~~~\textbf{$T_{Sf}$.setVariables($\mathit{DM}_{Sf}.getVariables()$)}	 

\State $T_{Sf}$.root = $root$

\State \textbf{Return} $T_{Sf}$

\end{algorithmic}

\end{algorithm}

The structure of the Stateflow model allows a multilevel hierarchy of states in which a state $S_{i.j}$ can contain sub-states with different types, where $i$ indicates the number of the level hierarchy of the Stateflow model ($ i= 0...n)$, $j$ is the number of states, and $n$ is the total number of levels in the Stateflow model. Therefore, the process of algorithm 2 traverses recursively the Stateflow data model based on the depth-first search algorithm to consider all sub-states of the superstate and add them to the tree of Stateflow. Each Stateflow model has two kinds of state decomposition: OR states (exclusive) and AND states (parallel) \cite{Math2016}. The Stateflow semantics allow every state to have a state decomposition that indicates what type of sub-states the superstate can contain. All sub-states of a superstate $S_{i.j}$ should have the same type of decomposition of the parent state.

\subsubsection{\textbf{Generating the tree of the Stateflow model}} The algorithm for generating the tree of the Stateflow (shown in algorithms 2 \& 3) starts by parsing the XML specifications of the Simulink's Stateflow $Sf$ into the data model $\mathit{DM}_{Sf}$ (see lines $1-3$). A tree Stateflow will be created to store a root node, a list of transitions and the list of the Stateflow variables. As a Stateflow model has no root state, a default node called $root$ will be created to store all information about the superstates at level 0 and assigned its \emph{ParentID} randomly as an integer number that is not assigned to any state in the Stateflow model (see lines $4-7$). Each node stores the following data: \emph{$id$}, \emph{name},  \emph{parentID}, \emph{$T$} a list of transitions, a list of children (sub-states), the order of execution, a list of the state actions (entry, during and exit actions) and type of decomposition state (\emph{OR  State} or \emph{AND State}). All superstates at level 0 in the Stateflow model are added as the children of the default \emph{root} node. For each state $S_{i,j}$, a node will be created to store all information of the state $S_{i,j}$ (see lines $8-14$). If the state $S_{i,j}$ has children, then all its sub-states will be traversed recursively until no more children exist for the superstate (see line $15$). Then, a state $node$ will be added as a child of the root node (see line $17$). The transitions at this level will be added to a transition list of the Stateflow tree $T_{sf}$ to be used in the next algorithms (see line $21-23$): the \emph {SMVGenerator} algorithm, Extended Finite State Machine model (\emph{EFSMGenerator}) and a truth-table of the EFSM model generator.

\begin{algorithm}[t]

\caption{traverseChildren(root, s)} 

\textbf{Input:} $root$ :  a  root node in the tree $T_{Sf}$, 
$s$: a state in a satateflow data model $\mathit{DM}_{Sf}$

\textbf{Description:}

\begin{algorithmic}[1]

\If{\textbf{$s$.hasChildren()==ture}}

\For {\textbf{ each State}  $child $ \textbf{in} $ s.getChildren()$}

\State\textbf{Create a new node $node$}

\State \textbf{Set} $node.setName$ $\leftarrow$ $child.getName$ 
\State  $nodel.setId$ $\leftarrow$ $child.getID$
\State $node.setParentID$ $\leftarrow$ $child.getParentID$
\State $node.setDecomposition$ $\leftarrow$  $child.getDecomposition()$ 
\State $node.StatesActions$ $\leftarrow$ $child.StatesActions()$//Entry, During, Exit Actions
\If{$child$.hasChildren()== ture}

\State $node.setHasChildren (true)$

\EndIf

\State~~~~\textbf{Add} $root$.addChild($node$)

\State ~~~~\textbf{traverseChildren(node, child)}

\EndFor

\EndIf

\end{algorithmic}

\end{algorithm}

 \begin{figure*}[t]
 	\centering
 	\includegraphics[width=3.5in]{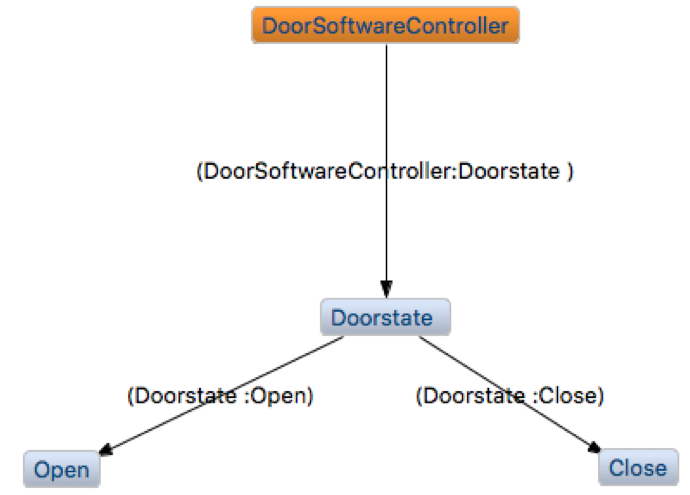}
 	\caption{The tree of the stateflow model of the train door software controller}
 	\label{tree}
 \end{figure*}

Figure \ref{tree} shows the tree of the stateflow model of the train door software controller.

\begin{figure} [t]
	
	\lstset{language=PHP,
		basicstyle=\ttfamily,
		keywordstyle=\color{blue}\ttfamily,
		stringstyle=\color{black}\ttfamily,
		commentstyle=\color{green}\ttfamily,
		breaklines=true
	}	
	\begin{lstlisting}[numbers=left]
MODULE main (<module variables>)
 VAR
 variables : <range data type>/<enumeration>
 <nameSub1>: _SubModule1 (variables) 
 ...
 <nameSubN>: _SubModuleN (variables) 
 states: <All children states>
 ASSIGN
  INIT (states=<initialState>)
  INIT (<variable> =<value>)
  next(<variable>):= case
  <var1>=<value> & <tranConditon>:<nextValue>;
  ...
  next(states):= case
  <states>=value & transition:nextState;
  ...
  esac;
 LTLSPEC
  <List of LTL formulae>
	\end{lstlisting}
	
	\caption{The structure of the SMV model is generated from the Stateflow tree and the STPA data object}
	\label{fig4}
\end{figure}

\subsubsection{\textbf{Generating the SMV model from the STPA and Stateflow data models}} Figure \ref{fig4} shows the basic structure of the $SMV$ model as described in \cite{Cavada2010}. Each $SMV$ module represents a superstate in the Stateflow model which can contain the following sections: 1) The name of the model with the optional state variable parameters, 2) The declaration of the state variable and their possible values, 3) The initial values of variables and the $states$ variable, 4) The sub-modules of the super module declaration, 5) The transitions of the module, and a list of the LTL formulae. To represent the states of the Stateflow model ($\simeq$ internal state variables of each controller in STPA) in an SMV model, we declare an enumeration variable called \emph{"states"} which contains the names of sub-states of the superstate in the Stateflow model. 

Based on the principles of the SMV model \cite{Cavada2010} and Stateflow diagram \cite{Math2016}, we develop an algorithm to transform the Stateflow (safe behavioural model) and STPA data objects into an SMV model. Algorithm 4 shows the process of automatically transforming the safe behavioural model and the STPA data models into an input language of the SMV model checker. The algorithm traverses the states of the safe behavioural model recursively and generates the SMV model by parsing the hierarchical levels of the safe behavioural model. The inputs of the algorithm are a tree of Stateflow model $T_{Sf}$ which is created based on algorithm 2 and the STPA data model $\mathit{DataModel_{SW}}$ of the software controller $\mathit{CO}_{i}$, which is generated based on algorithm 1 and a  node $n$ in the tree $T_{Sf}$. 

The algorithm $4$$-$$5$ process starts by creating an object of the $SMV$ model which represents all structure data of the $SMV$ model (see line $1$). The algorithm takes the root node of the safe behavioural model tree as the input at the first time to create the main module of the SMV model, then it cerate the main module section (see lines $2-6$)  and declares the \emph {VAR} section (see line $7$). In this section, the algorithm will declare the local variables of the root node and maps their data types to the $SMV$ data types (see lines $8-9$). The algorithm will check whether the variables are declared exactly in the process model of the software controller in the STPA control structure diagram to reduce the time and effort of matching these variables during the verification step (see line $10$). If a name of variable or state in the STPA data model does not match any name in the data model of the safe behavioural model, then the algorithm will show a message to the user and return \emph{null} (see line $49$).  

The SMV model does not support the same basic data types (int, double, single) as the data types which are declared in the Stateflow model, it supports only a finite range type as integer range $min...max$ value. Therefore, the algorithm should map the data types (int, double or single) into a finite range which starts with a minimum value and ends with a maximum value of integer data type. The enumeration data types are declared into the Stateflow model as a class which is saved in a separate file and not in the XML specifications of Stateflow model. Therefore, the algorithm checks each variable with enumeration data type whether it is a process model variable in the STPA data model or not. In case the enumeration variable is a process model variable, the algorithm takes its values as they are defined in the STPA process model variable values. Otherwise, the algorithm creates an empty bracket $\{ \}$ for the values of the enumeration variable and prompts the user to determine the values of this variable (see lines $11-15$). 

Next, the algorithm checks whether the root state $root$ has children states and which of them has children too (see line $16-20$). In case that a child $node$ of $root$ has children, then the algorithm declares a sub-module for this child $node$. Then, the algorithm takes all variables of the current state $node$ to create a list of the parameters of the sub- module (see line $18$). Next, the algorithm parses the sub-states of the superstate and creates the variable \emph{"states"} with a list of the names of the sub-states as values (see lines $21-24$). 
\begin{algorithm}[p]
	\caption{generateSMV($T_{Sf}$, $\mathit{DataModel}_{SW}$, $n$)} 
	\textbf{Input:} $T_{Sf}$ :  a  tree data model of safe behavioural model,
	$\mathit{DataModel}_{SW}$: a STPA data model of controller $CO_{i}$, $n$: is a node in tree $T_{Sf}$.\\
	\textbf{Output:} $\mathit{SMV}_{i}$: an SMV object represents the data of SMV model.\\
	\textbf{Description:}
	\begin{algorithmic}[1]
		\State ~~~\textbf{Create} a $\mathit{SMV}_{i}$ $\leftarrow$ SMV model object
		\If { ($n$.isRoot ()==true)} 
	\State  ~~~\textbf{Set} header of $\mathit{SMV}_{i}$ $\leftarrow$ \textbf{'Module main'}
		\Else 
	\State $\mathit{SMV}_{i}$$\leftarrow$ \textbf{'Module'} root.getName() (root.getVariables)
		\EndIf
		\State\textbf{Set} \textbf{VAR} section of $\mathit{SMV}_{i}$ $\leftarrow$ \textbf{'VAR'}
	\State \textbf{Parse Variables}  $\mathit{SMV}_{i}.setVariables ()$ $\leftarrow$ $T_{Sf}.getVariables()$
	\State\textbf{Map} data type of $SMV$ variables into SMV data types. 
		\If { (ValidateSTPADataModel ($n.getVaraibles()$, $\mathit{DataModel}_{Sw}$)} 
		\If {($v.getType ()!="Enum"$)}
	\State\textbf{Declare} each variable as $v.getName:$ $v$.getType(); 
		\Else  \State	$v.getName: \{  \}$; // an empty bracket 
			\EndIf	 
		\If {($n$.isSubModule==true)}
		\For {\textbf{each} s $\in$ $n$.getChildren()}
	\State \textbf{Declare} "Sub\_" + s.getName($n.getVariables ()$) ($n.getVariableNames()$)
		\EndFor
		\EndIf
	\State \textbf{Declare} $states$ variable in  $SMV$ $\leftarrow$ \textbf{'states'}
			\For {\textbf{each} s $\in$ $n$.getChildren()}
	\State  states $\leftarrow$ $.s.getName()$ 
		\EndFor
	\State \textbf{Set ASSIGN} section of $SMV$ $\leftarrow$ \textbf{'ASSIGN'}
		\State\textbf{initial} each $v$ of $\mathit{SMV}_{i}$ $\leftarrow$ 
		\State init($v$.getName()) :=initial\_Value;
	\State  \textbf{Parse} Transitions $T$ $\leftarrow$ $n$.getTransitions()    
	\State\textbf{Set} Next section of  $T$ of $n$ state  
		\State $\mathit{SMV}_{i}$.$\leftarrow$ \textbf{'next'} (states) := case 
		\For { \textbf{each} t $\in$ $T$ }
	\State $states:=t.Source \& t.Condition: t.Destination;$
	\State  TRUE: states ; esac 
		\EndFor
		\If { ($n$.isRoot ()==true)} 
		\For {\textbf{each} v $\in$ $n$.getVariables ()}
	\State $\mathit{SMV}_{i}$.$\leftarrow$ \textbf{'next'} (v.getName()) := case 
		\State states=n.getSource(): n.getEntryDuringExit(v.getFunction()) 
	\State TRUE: v.getName (); esac;
	\EndFor 
		\EndIf
		
			\algstore{myalg}
				\end{algorithmic}
		\end{algorithm}
\begin{algorithm}[t]
	\caption{generateSMV($T_{Sf}$, $\mathit{DataModel}_{SW}$, $n$) (continued)}
	\begin{algorithmic}
		\algrestore{myalg}

	\State$\mathit{SMV}_{i}$ $\leftarrow$ \textbf{'esac;'}
		\If {($n$.hasChildren())}
		\For {s $\in$ $root$.getChildren() }
\State $\mathit{SMV}_{i}$ $\leftarrow$ generateSMV($T_{Sf}$, $\mathit{DCs}_{i}$, $s$)
		\EndFor
		\EndIf   
		\Else \State \textbf{Show} "STPA variables do not match $Sf$  variables" \&  \textbf{Set}  $\mathit{SMV}_{i}$ $\leftarrow$  null
		\EndIf
	\State $\mathit{SMV}_{i}$$\leftarrow$ "LTLSPEC "$\mathit{DCs}_{i}$.getLTL()
		\State\textbf{Return} $\mathit{SMV}_{i}$.
		
	\end{algorithmic}
\end{algorithm}
The algorithm will create the seciton \emph{"Assign"} to initial the states and variables of the SVM model (see line $15$). The algorithm will create the \emph{initial} expression of the \emph{"states"} variable. Each data variable will also be initialised with the minimum value of its data type such as a variable with a numeric data type with zero, Boolean with FALSE and enumeration variable with the first value (see in lines $26-27$). Next, the algorithm will parse all transitions of the current state $node$ and create the \emph{next} expressions for the \emph{"states"} variable (see lines $28$$-$$34$). The \emph{next} expressions of $states$ variable refer to the transition relations of current state \emph{$node$} with other states in the model (the truth-table). The \emph{next} expressions of the \emph{$states$} variable are expressed as follows:  
\begin{lstlisting}
next(states):= case
states=<sub-state> : <nextstate>
...
1: {All sub-states}; esac;

\end{lstlisting}
To create the \emph{next} expressions for each data variable, the algorithm parses the \emph{Entry, During and Exit} actions of the current state and extracts all actions of each variable (see lines $35-41$). The \emph{next} expressions of the data variables refer to the values of variables in the next state. The \emph{next} expressions of each data variable are expressed as follows: \\
\begin{lstlisting}
next(variable):= case
states = <state> & transition: <nextValue>
....
\end{lstlisting}

The algorithm will continue parsing the superstate in the tree of the safe behavioural model (Stateflow) till all superstates have been visited (see lines 43--48). The generated SMV specifications of each sub-module and the main module will be saved as a string into a stack object. Finally, the algorithm will fetch the LTL formulae from the STPA data model object and add them at the end of the main-module section (see lines 50--51). 

To check the correctness of the generated SMV model and the safe behavioural model, we run the NuSMV model checker to verify whether the SMV model contains errors and verify it against the STPA software safety requirements expressed in the LTL formulae and saved to the SMV model.  

Figure \ref{fig44} shows an example of the generated SMV model of the safe behavourial model of the train door software controller.

 \begin{figure} [t]
 	
 	\lstset{language=PHP,
 		basicstyle=\ttfamily,
 		keywordstyle=\color{blue}\ttfamily,
 		stringstyle=\color{black}\ttfamily,
 		commentstyle=\color{green}\ttfamily,
 		breaklines=true
 	}	
 	\begin{lstlisting}[numbers=left]
  MODULE Sub_Doorstate (Trainposition,Trainstatus,PersonIndoorway)
  VAR
   controlAction:{Opendoor,Closedoor,Stop};
   states: {Open,Close};
  ASSIGN
   init (states):=Open;
  next (states):=case
  states=Open & (Trainposition=Aligend & Trainstatus=Stop):Open;
  states=Close & (PersonIndoorway) : Open;
  states=Open & (!PersonIndoorway) : Close;
  states=Close & (Trainposition=Aligend & Trainstatus=Stop) : Close;
  TRUE: {Open ,Close };
  esac;
  
  MODULE main
  VAR
  PersonIndoorway: boolean; 
  Trainposition: {Aligned, NotAligned}
  Trainstatus: {Stop, Move}
  Doorstate :Sub_Doorstate (Trainposition,Trainstatus,PersonIndoorway);
  states: {Doorstate};
  ASSIGN
  init (states):=Doorstate ;
  init (PersonIndoorway) := FALSE ;
  next (states):=case
  TRUE:{Doorstate};
  esac;
  LTLSPEC G (((trainstatus== stop) & (doorstate == close) & (trainposition== Aligned)  & (PersonIndoorway==TRUE))  -> !(controlAction== closedoor))  
  
   	\end{lstlisting}
 	
 	\caption{An example of the generated SMV model for the train door software controller}
 	\label{fig44}
 \end{figure}

\subsection {Automatically Constructing the Safe Test Model from the Safe Software Behavioural Model }

After ensuring the correctness of the generated SMV model of the safe behavioural model (Stateflow model), the safe behavioural model which uses the notations of the Simulink's Stateflow should be transformed into the EFSM notation. For this purpose, we develop an algorithm to map the Stateflow tree of the safe behavioural model and its truth-table into an EFSM model. The algorithm 6--7 shows the process of transforming the tree of the Stateflow model into an EFSM model. The idea here is to eliminate the hierarchical and concurrent structure of the Stateflow model (flattened and broadcast communication) and transform them into the EFSM notations by considering the state decomposition (exclusive or parallel).

\begin{algorithm}[ht!]
	\caption{GenerateEFSM ($T_{Sf}$)} 
	\textbf{Input:} $T_{Sf}$ :  a  tree of Stateflow model,\\ 
	\textbf{Output:} $EFSM$: a Java object represent all data of EFSM\\
	\textbf{Description:}
	\begin{algorithmic}[1]
		\State \textbf {Create} StateNode $root$  $\leftarrow$ $T_{Sf}$.getRoot()
		\State \textbf{Get} TruthTable $truthTable$ $\leftarrow$ $T_{Sf}$.getTruthTable()
		\If{\textbf{$root$.hasChildren()==ture}}
		\State \textbf{Set} Initial state $\leftarrow$  $T_{Sf}$.getInitialState( )
		\While {isHasSuperState($truthTable$) }
		\For { Transition $t$ $\in$ $truthTable$}
		\State \textbf{StateNode} src  $\leftarrow$ $t$.getSourceNode ()
		\State \textbf{StateNode} dest  $\leftarrow$ $t$.getDestinationNode ()
		
		\If{\textbf{$src$.isSuper() \& !($dest$.isSuper())}}
		\State \textbf{get} $children$ $\leftarrow$ $src$.getChildren() 
		\For { $child$ $\in$ $children$}
		\State updateTruthTable ($child, dest, t, truthTable$)
		\EndFor
		
		\Else \If{\textbf{!($src$.isSuper()) \& $dest$.isSuper()  }} 
		\If {$dest$.Decomp('AND\_STATE') }
		\State get children $\leftarrow$ $dest$.getSubSates();
		\For { $child$ $\in$ $children$}
		\State updateTruthTable (src, child, t, truthTable)
		\EndFor
		\Else \If {$dest$.Decomp('OR\_STATE')}
		\State $S_{D} $$\leftarrow$ getDefaultState($dest$)
		\State updateTruthTable (src, $S_{D}$, t, truthTable)
		\EndIf
		\EndIf
		\EndIf
		\If{\textbf{$src$.isSuper() \& $dest$.isSuper() \&   $dest$.Decomp('OR\_STATE')}} 
		\State \textbf{get} srcChildren $\leftarrow$ $src$.getSubSates();
		\State \textbf{get} def $\leftarrow$ $dest$.getDefaultState();
		\For { $s$ $\in$ $srcchildren$}
		
		\State updateTruthTable (s, def, t, truthTable)
		
		\EndFor
		\EndIf

		\algstore{myalg}
	\end{algorithmic}
\end{algorithm}   

\begin{algorithm}[t]
	\caption{GenerateEFSM ($T_{Sf}$) (continued)}
	\begin{algorithmic}
		\algrestore{myalg}

		\If{\textbf{$src$.isSuper() \& $dest$.isSuper() \&  $dest$.Decomp('AND\_STATE')}  } 
		\State \textbf{get} srcChildren $\leftarrow$ $src$.getSubSates();
		\State \textbf{get} destChildren $\leftarrow$ $dest$.getSubSates();
		\For { $s$ $\in$ $srcchildren$}
		\For { $d$ $\in$ $destchildren$}
		\State updateTruthTable (s, d, t, truthTable)
		\EndFor
		\EndFor
		\Else \If{\textbf{!($src$.isSuper())\& !($dest$.isSuper())}} 
		\State updateTruthTable (src, dest, t, truthTable)
		\EndIf
		\EndIf
		\EndIf
		
		\EndFor
		\EndWhile
		
		\EndIf
		\State  \textbf{Add} $EFSM$.setTruthTable $\leftarrow$  truthTable
		\State  \textbf{Add} $EFSM$.setStates $\leftarrow$  $T_{Sf}$.getStates()
		\State \textbf{Return} $EFSM$.
		
	\end{algorithmic}
\end{algorithm}

The algorithm 6--7 starts by taking the root node of the Stateflow tree $T_{Sf}$ as the root node of the EFSM model and the truth-table of the Stateflow as the truth-table of the EFSM model (see line 1). The Stateflow semantic supports multi-hierarchy levels of states, whereas the EFSM model does not. Therefore, the truth-table of the EFSM model must not have any source or destination node as a superstate (a state that has children). The idea here is to investigate the truth-table of Stateflow and update the destination and source parent state with its sub-states. At the beginning, the algorithm checks whether there is a superstate in the truth-table (see lines $2-3$). For each transition $t \in T$ in the truth-table, the algorithm will identify its source and destination states and create two state nodes (see lines 6--8). Next, the algorithm will check their state decomposition as follows:  

\begin{itemize}
	\item If source state $src \in T_{sf}$ of transition $t$ is a \textbf{superstate} with a state decomposition ``OR\_STATE" or ``AND\_STATE" and the destination node $dest \in T_{sf}$ is \textbf{not superstate}. Each sub-state of $src$ state must be linked to the destination state $dest$ by creating a new transition with the same information of transition $T\in T_{sf}.TruthTable$ for each sub-state and only update the source with sub-state (see lines $9-14$).
	
	\item If source state $src \in T_{sf}$ is \textbf{not superstate} and the destination state $dest \in T_{sf}$ is \textbf{superstate} with a state decomposition ``AND\_STATE". All sub-states of $dest$ state should be identified and linked with the source state (see line $15-21$). Algorithm 7 will create a new transition for each sub-state of $dest$, where   source is $src$ and destination is the sub-state of destination. 
	
	\item If source state $src \in T_{sf}$ is \textbf{not superstate} and the destination state $dest \in T_{sf}$ is \textbf{superstate} with a state decomposition ``OR\_STATE". The default state $defaultState$ of superstate $dest$ (a default state is a state which has a default transition) should be identified (see lines $22-27$). Algorithm 7 will create a new transition and set its source as $src$ and its destination as the default state of destination.
	
	\item If source state $src \in T_{sf}$ is \textbf{superstate} with a state decomposition ``OR\_STATE" or ``AND\_STATE"  and the destination state $dest \in T_{sf}$ is \textbf{superstate} with a state decomposition ``OR\_STATE". All sub-states of $src$ state should be identified and linked with a default state of $dest$ state (see lines $28-34$). Algorithm 7 will create a new transition for each sub-state of $src$ and its source is $src$ and its destination is the default state of destination $dest$ state. 
	
	\item If source state $src \in T_{sf}$ is \textbf{superstate} with a state decomposition ``OR\_STATE" or ``AND\_STATE"  and the destination state $dest \in T_{sf}$ is \textbf{superstate} with a state decomposition ``AND\_STATE". All sub-states of $src$ state should be identified and linked with all sub-states of  $dest$ state (see lines $35-43$). Algorithm 7 will create a new transition for each sub-state of $src$ and its source is $src$ and its destination is the sub-state of destination $dest$ state. 
	
	\item If source state $src \in T_{sf}$ is \textbf{not superstate} and the destination state $dest \in T_{sf}$ is \textbf{not superstate}. A transition $t$ will be added into the truth-table (see lines $44-46$).  
	
\end{itemize}

The algorithm runs continuously till no superstate exist in the truth-table. All sub-states (without children) in the Stateflow model tree will be taken as the states of the EFSM model. Also, all data variables of the Stateflow model and the actions of the state (entry, exist, during) will be added into the states of EFSM.

\begin{algorithm}[t]
	\caption{UpdateTruthTable ($src$, $dest$, $t$, $truthTable$)} 
	\textbf{Input:} $src$ : a source node of transition $t$, $dest$: a destination node of transition $t$, $t$ : a transition in the truth-table, $truthTable$: a truthTable of Stateflow tree $T_{sf}$\\ 
	\textbf{Description:}
	\begin{algorithmic}[1]
		\State \textbf {create} new Transition $t\_new$
		\State \textbf {set} data $t\_new$ $\leftarrow$ $t$
		\State \textbf {update} $t\_new$.setSrc(src)
		\State \textbf {update} $t\_new$.setDest(dest)
		\State \textbf {add} $truthTable$ $\leftarrow$ $t\_new$
		
	\end{algorithmic}
\end{algorithm}

 \begin{figure*}[t]
 	\centering
 	\includegraphics[width=5.0in]{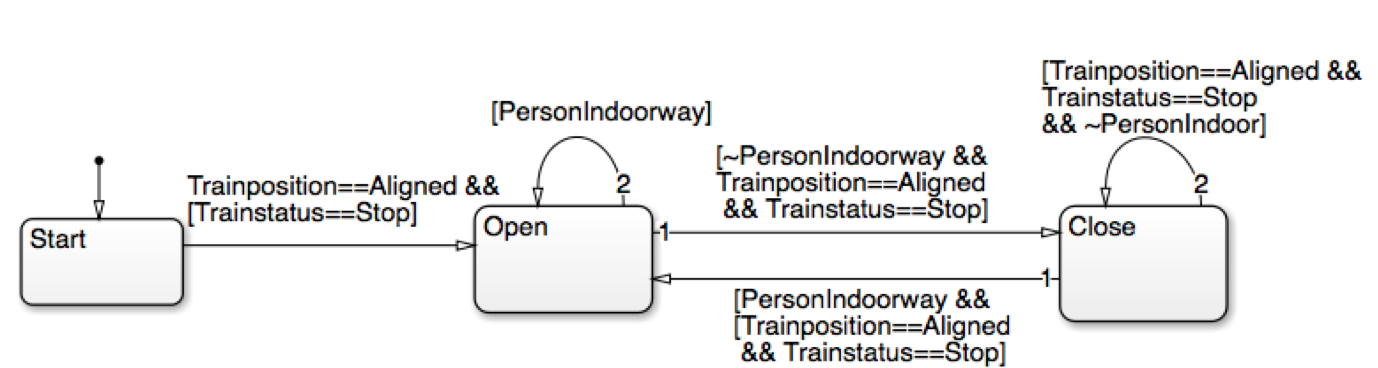}
 	\caption{The safe test model (EFSM) of the train door software controller}
 	\label{EFMS}
 \end{figure*}

Figure \ref{EFMS} shows an example of the generated EFSM of the train door software controller

\subsection {Automatically Generating Safety-Based Test Cases}
The final step is to generate the test cases from the safe test model (extended finite state machine) which are constructed from the safe behavioural model.

We developed a random walk-based algorithm for automatic test case generation from the safe test model. We implemented three search-based algorithms (e.g. depth-first search, breadth-first search, and both combined depth-breadth-first search). The idea behind here is to select a state in the safe test model as a start node and transform into a Java Script function at run time. The Java script function takes the variables which are declared in the state actions (Entry, during, Exit) of each as parameters and executes the state actions to update the values of the variables. The return value of the function will be determined based on the data type of each variable which is declared in the Simulink Stateflow model. Next, the algorithm will check the transition conditions of a state to determine which is the next state. During traversing the safe test model, the information of the visited states (path sequences) will be saved in a test suite.

Generating test cases from a model usually leads to an infinite number of possible test cases. Therefore, it is necessary to choose a suitable test coverage criteria to manage the generating process. In our algorithm, we identify three test coverage criteria: 1) \emph{state coverage} which is the number of visited states divided by the total number of the states of the model, 2) \emph{transition coverage} is the number of the executed transitions divided by the total number of the transitions, 3) \emph{STPA safety requirements coverage} in which each STPA software safety requirement should be covered at least in one test case to trace how the STPA-generated software safety requirements are covered into the generated test cases. To measure the STPA SSR coverage, we define a \emph{safety requirements traceability} matrix between the generated safe test model and STPA software safety requirements to manage the quality of the test case generating process and measure the coverage of STPA software safety requirements in the generated safety-based test cases. As the safe test model of the safe behavioural model is constrained with STPA safety requirements (step 2) and contains the process model variables as states, the algorithm will automatically generate the traceability matrix ($TM= SSR \times TN$, where $\mathit{SSR}$ $\in$ $DCs$ of the STPA data model and $TN$ transition conditions $\in$ $T_{Sf}$). 

\begin{algorithm}[t]
	\caption{Generate Traceability Matrix ($SSR$, $TN$, $src$, $minSimilarity$)} 
	\textbf{Input:} $SSR$: a STPA-generated software safety requirement,  
	$TN$: a transition condition in a safe test model extracted from $SBM$. 
	$src$: a source node of transition condition $Tn$.\\
	$minSimilarity$: a minimum degree of similarity between 5\% ... 100\%. \\
	\textbf{Output} $TM$: a traceability matrix.
	\label{alg:5.60} 
	\textbf{Description:}
	\begin{algorithmic}[1]
		\State \textbf {Add} $TN$ $\leftarrow$ $states=src.getName()$
		\State \textbf {Add} $TN$ $\leftarrow$ $controlAction=src.getAction().getName()$
		\State \textbf {tokenize} $SSR[]$  $\leftarrow$ $SSR$
		\State \textbf {tokenize} $TN[]$  $\leftarrow$ $TN$
		\State \textbf {get} $max\_Tokens$  $\leftarrow$ max ($SSR[ ]$, $TN[ ]$)
		\State \text{inital} $Sim$  $\leftarrow$ $0$
		\State \text{inital} $matched\_Tokens$  $\leftarrow$ $0$
		\State \text{inital} $i$  $\leftarrow$ $0$
		
		\While {$i$ $<$ $max\_Tokens$}
		\State \text{inital} $j$  $\leftarrow$ $0$
		\While {$j$ $<$ $max\_Tokens-1$}
		\If {$SSR[i]== TN[j]$}
		\State   $matched\_Tokens$ = $matched\_Tokens +1$
		\EndIf
		\State	$j = j +1$ 
		\EndWhile
		\State	$i = i +1$ 
		\EndWhile
		\State $Sim_{SSR, TN}$ =($matched\_Tokens$ / $max\_Tokens$) $\times$ 100
		
		\If {$Sim_{SSR, TN}$ > $minSimilarity\%$}
		\State	 \textbf{Add} $TM$ $\leftarrow$ $SSR$ $\times$ $TN$
		
		\EndIf
		
		\State \textbf{Return} $TM$.
	\end{algorithmic}
\end{algorithm}

Algorithm 9 shows how to generate the traceability matrix $TM$ by calculating the similarity degree between each STPA-generated software safety requirement($SSR$) and the transitions condition ($TN$) of the safe behavioural model and the input state actions of the source state of transition condition $TN$.  The similarity degree is calculated by the following equation:
\begin{equation} \label {eq1}
\mbox{$Sim_{(SSR,TN)}$ } = \frac{|\mbox{\#Total No. matched tokens between (SSR, TN)}|}{| \mbox{\#Max No. tokens in (SSR, TN) }| } \times 100
\end{equation}

\begin{figure*}[t]
	
	\centering
	
	\includegraphics[width=4.5in]{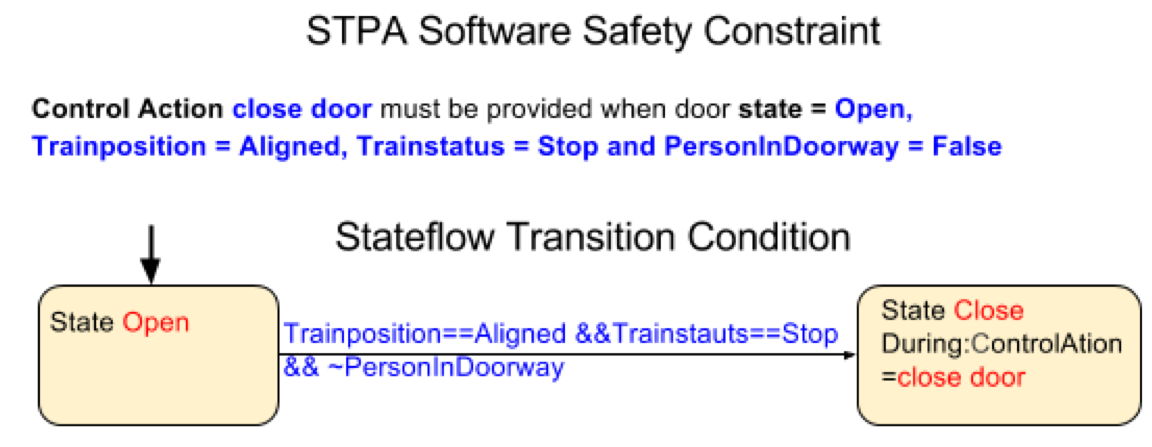}
	
	\caption{An example of the similarity degree between STPA safety requirements and Stateflow transition conditions}
	\label{similarity}
\end{figure*}

Algorithm 9 takes a STPA-generated software safety requirement ($SSR$), a transition condition $TN$, the source state of the transition condition $TN$ and a minimum degree of similarity ($maxSimilarity$) which should be between 5\% \ldots 100\% and entered by the user. To compare between the STPA-generated software safety requirements and transition conditions, the algorithm construct at first the full transition information by including the name of source state and the control action which is provided in this state to the transition condition (shown in Fig. \ref{similarity}). The algorithm creates the full transition information by adding the source state $src$ and the control action $controlAction$ to the transition condition $TN$. The full transition condition will be constructed as follows:

\emph{$full transition$  $\leftarrow $ \{states==src.getName() and contorlAction==src.getAction() and src.getTransitionCondition (TN)\}}

The algorithm calculates the similarity degree based on the equation \ref{eq1}. If the similarity degree is greater than the minimum degree of similarity, then the algorithm will create an item in the traceability matrix for the software safety requirement $SSR$ and the transition condition $TN$. The algorithm also allows the user to set the maximum similarity degree between $5..100\%$ before generating the safety-based test cases.

\begin{algorithm}[t]
	\caption{Generate Safety-based Test Cases($STM$, $TM$,$CC$, $TestSteps$,  $StopConiditon$)} 
	\textbf{Input:} $STM$: a safe test model extracted from $SBM$, $TM$: a traceability matrix, $CC$: is a list of the test coverage criteria, $TestSteps$ is the total number of execution algorithms, $StopCondition$: a condition to stop the execution process.\\ 
	\textbf{Output} $TS$: a list of test suites, each test suite should contain a list one test case $TC$.
	
	\textbf{Description:}
	\begin{algorithmic}[1]
		\State \textbf {Initial} step  $\leftarrow$ 0
		\While {step $<$ TestSteps}
		\State \textbf {Choose} $start$ state $\leftarrow$ $STM.getRandomState()$ 
		\State \textbf {Choose} $end$ state $\leftarrow$ $STM.getRandomState()$ 
		\State \textbf {Create}  a new test suite $ts$
		\If {$StopConiditon<100.0\%$}
		\State \textbf {Randomly} Generate\_Tes\_InputData ()
		\State \textbf {Walk} $\mathit{TC}_{i}$ $\leftarrow$ GenerateTestCasesByDFS (start, end)
		\State \textbf {Add} $ts$ $\leftarrow$ $\mathit{TC}_{i}$ 
		\State \textbf {Walk} $\mathit{TC}_{j}$ $\leftarrow$ GenerateTestCasesByBFS (start)
		\State \textbf {Add} $ts$ $\leftarrow$ $\mathit{TC}_{j}$ 
		\Else
		\If {$StopConiditon$==$100.0\%$} 
		\State	\textbf {Calculate\_Coverage\_Criteria()}
		\State	\textbf {STOP}
		\EndIf
		\EndIf
		\State {ADD} $TS$  $\leftarrow$ $ts$
		\State	Calculate\_Coverage\_Criteria()
		\State unvisitedTransitions(STM) 
		\State unvisitedStates(STM) 
		\State \textbf {Initial} step  $\leftarrow$ step + 1
		
		\EndWhile
		\State \textbf{Return} $TS$.
	\end{algorithmic}
\end{algorithm}

Algorithm 10 shows how to generate the test cases from the safe test model. It takes the generated Safe Test Model ($\mathit{STM}$), a Traceability Matrix $TM$, a list of the test coverage criteria $\mathit{CC}$, a number of test steps which is the total number of executions of the algorithm and a stop condition which is a test coverage criteria to stop the execution of the algorithm when it reaches 100\%. The process of generating the test cases from the safe test model can be described as follows: 

\begin{enumerate}
 
	\item The algorithm starts by selecting a random state as the start state and a state as the end state from the safe test model to generate all possible paths between them (see lines $3-4$).
	\item A new test suite $ts$ will be created to store all the generated test cases. 
	\item Generate for each input data variable a random value between its minimum and maximum values which are identified by the user (see line $7$). 
	\item Walk randomly by using the depth-first algorithm, all possible paths between the start and end states will be identified. The path here means a sequence of the visited states and their transitions. We also use the breadth-first algorithm combined with depth-first algorithm to identify all possible paths $PT$ from start state to achieve a good test coverage criteria (see line $8-11$). 
	\begin{itemize}
		\item For each transition $t$ in path $pt \in PT$, its transition condition will be transformed into a Java Script function. The test input variables $in$ will be passed as an input of a Java Script function. To execute this function at the run time, we use the Java Script Engine which invokes the function with values of input data parameters and returns the result.
		\item For each state $s$ in path $pt$, the state actions (Entry, During, Exit) will be eliminated and transformed into Java Script functions. These will be executed to update the values of each local $loc$ or output variable $out$ of each state.     
		\item Create a new test case $tc$. Each test case will store the information about the sequence path $pt$ such as: \emph{id} is a number of the test case, \emph{id\_Ts} which is the number of the test suite, \emph{id\_SSR} which is the number of the software safety requirement, \emph{preconditions and actions} which is the sequence of the local variables of states in the path $pt$ and their updated values, and \emph{postconditions} which is the sequence of output variables and their values. 
		
	\end{itemize} 
	\item Check whether the test case $tc$ has been covered in any test suite. If it hasn't, $tc$ will be added to the test suite $ts$ (see line $13$).
	\item Calculate the test coverage criteria and check the stop condition of the algorithm (see line $14$). 
	\item Change status of all states and transitions in the safe test model to unvisited to generate a new sequence path (see line $20-21$).  
	\item The algorithm will be continued (repeat 1-8) till the stop condition is achieved (100\%) or the number of executions the algorithm has been reached to the total number of the test steps. 
\end{enumerate}

Ultimately, the time spent during test case generation process, the values of the test coverage criteria and a list of test suits and their test cases with the related software safety requirements will be automatically saved into a CSV file.

\section{Tool Support}
\begin{figure*}[t]
	\centering
	\includegraphics[   width=14cm, height=7cm]{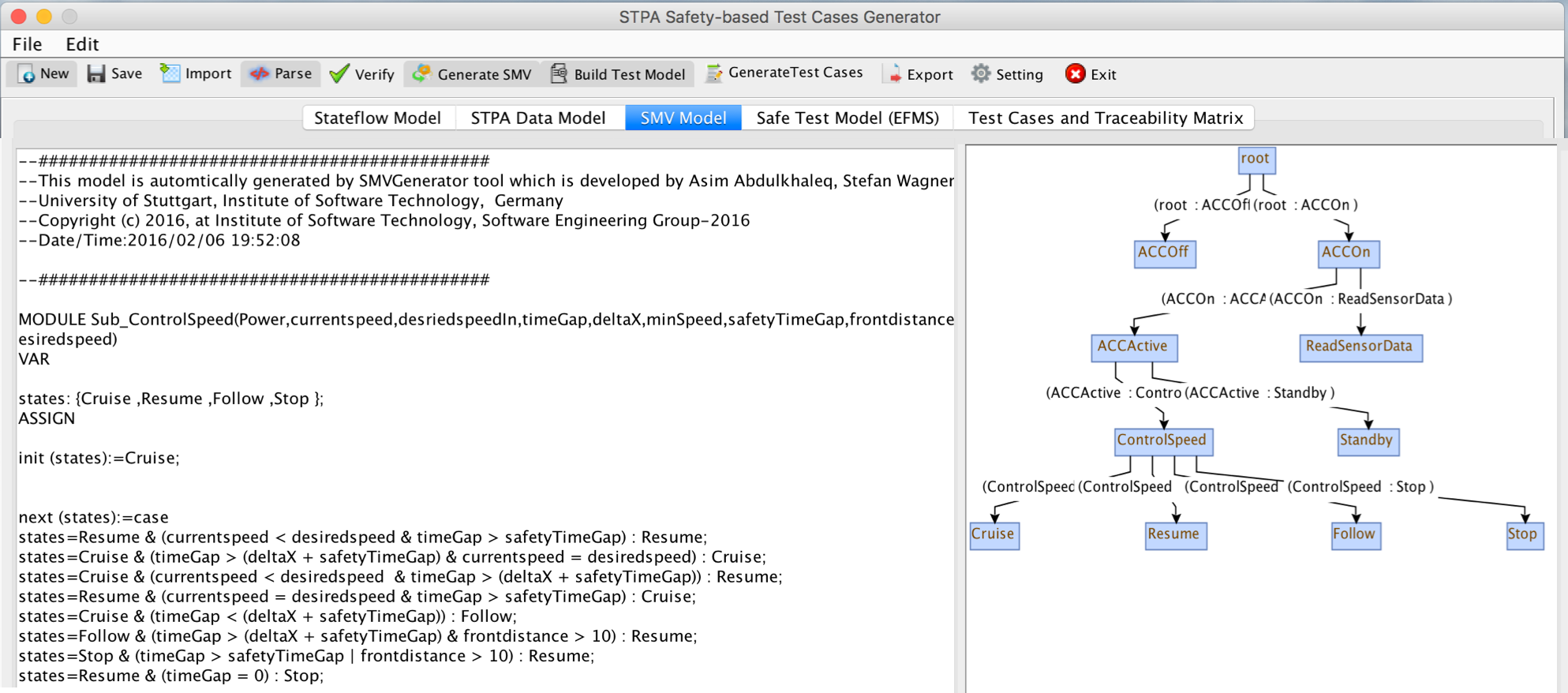}
	\caption{STPA TCGenerator: STPA test cases generator tool}
	\label{figSTPATC}
\end{figure*}
Here we describe the implementation of the proposed approach for generating test cases based on the information derived from the STPA safety analysis. We use the previous algorithms and rules as the basis for implementing tool support for our safety-based test case generation approach. 

To automatically formalize the STPA software safety requirements which are documented in XSTAMPP and transformed into LTL based on rules $1$$-$$4$, we developed an Eclipse plug-in called XSTPA\footnote{\url{http://www.iste.uni-stuttgart.de/se/werkzeuge/xstpa.html}} based on the XSTAMPP architecture.  XSTPA automatically generates the context tables (combinations between process model variables) by using a Java library for the combinatorial testing algorithm called ACTS\footnote{\url{http://csrc.nist.gov/groups/SNS/acts/index.html}} \cite{kuhn2013introduction} which was developed by the American National Institute of Standards and Technology to generate combination sets of \emph{t} parameters with \emph{n} values. Based on rules 3 and 4, XSTPA automatically generates the LTL formulae. The generated LTL formulae will be used to check the correctness of the constructed  safe test model which will be used to generate the safety-based test cases for each STPA-generated software safety requirement.  

To generate the test cases based on STPA results, we implemented a tool support called \emph{STPA Test Cases Generator} (STPA TCGenerator, shown in Fig. \ref{figSTPATC}) which parses the STPA file project created in XSTAMPP and the safe behavourial model which is created with Simulink's Stateflow editor to generate the SMV model and check the correctness of the safe beahvioural model, eliminate the safe test model and generate safety-based test cases. In the following, we summarise the main functions of the STPA TCGenerator:

\begin{itemize}
	
	\item Parse the STPA data model which is documented in XML specification into Java objects.
	\item Parse the XML specification of the Simulink Stateflow model into Java objects.
	\item Based on the STPA data model and the Simulink Stateflow model, the tool automatically generates the SVM model.
	\item Check the consistency between the STPA data model and the specification of Simulink Stateflow and provides the results to the user (e.g. matched, does not match, and unknown).
	\item Verify the generated SMV model against the generated LTL of the STPA safety requirements. 
	\item Transform the Simulink Stateflow model into the extended finite state model for testing purposes.
	\item Generate the tractability matrix between STPA safety requirements and the Simulink Stateflow specifications.
	\item Allow the user to enter the test input data for each input variable. 
	\item Allow the user to configure the test case generation process by adding a number of test steps and selecting the test case generation algorithm and the test coverages.
	
\end{itemize}

To support software and safety engineers who use the XSTAMPP platform, we developed an Eclipse plug-in for the STPA TCGenerator tool called \emph{STPA TCGeneratorPlugin} which is integrated into the XSTAMPP platform to generate the test cases for each STPA software safety requirement within the XSTAMPP platform.

The first prototype of the \emph{STPA TCGenerator} standalone version and the results of the illustrative example are available online in our repository\footnote{\url{https://sourceforge.net/projects/stpastgenerator/}.}. The updatesite of the STPA TCGeneratorPlugin are available online in our repository \footnote{https://sourceforge.net/projects/stpatcgeneratorplugin/}.

 \section {An Illustrative Example: A Simulator of the ACC System with STOP and GO Function}

 To illustrate the proposed approach, we developed a simulator software written in ANSI-C to simulate the adaptive cruise control system with stop and go function by using two LEGO EV3 Mindstorm robots\footnote{\url{http://www.iste.uni-stuttgart.de/se/forschung/werkzeuge/acc-simulator/}}. The simulator was developed by a bachelor student within 6 months. The ACC with stop and go function \cite {stopendgo} is an extended version of the normal adaptive cruise control system. It maintains a certain speed and keeps a safe distance from the vehicle ahead based on the radar sensors. The ACC with stop and go function will bring the vehicle to a complete stop when the vehicle ahead comes to a standstill or there is a stationary object in the lane.

 Figure \ref{figACCmechanism} shows the mechanism of the simulator of the ACC with stop-and-go function. The ACC simulator maintains a constant time gap to vehicles ahead. It uses a forward ultrasonic sensor with a range of up to 255 centimeters, which is located in the front of the robot to detect the distance of the robot ahead of it and can automatically maintain the pre-set time gap. It adjusts the robot speed by increasing or decreasing the value of current speed to keep a safe distance. If the robot ahead is completely stopped, then the ACC simulator will slow down the robot vehicle to a standstill. If the vehicle ahead starts moving again, then the ACC simulator will automatically start to move again and maintain a constant time gap between the robot ahead. Our simulator algorithm is the ACC simulator starts first read the distance data from the ultrasonic sensor and then computes the time gap by using the following equation: 
 
  \begin{figure}[t]
  	\centering
  	\includegraphics[width=3in]{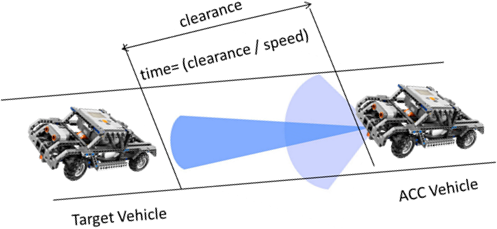}
  	\caption{A mechanism of the simulator of ACC with Stop and Go function}
  	\label{figACCmechanism}
  \end{figure}
  
  \begin{figure}[t]
  	\centering
  	\includegraphics[width=3.0in]{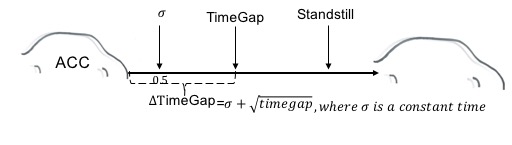}
  	\caption{The ACC system with a Stop and Go function scenarios}
  	\label{ACCScenarios}
  \end{figure}

 \begin{equation} 
 currentTimegap  =   |\frac{Frontdistance}{CurrentSpeed} | 
 \end{equation}
 Second, the simulator computes the standstill time, which is the time at which the ACC vehicle must decrease the speed or stop when the vehicle ahead is close or fully stopped. It is calculated as   
 \begin{equation} 
 \Delta Timegap =  stillstandtime  + \sqrt{currentTimegap} 
 \end{equation}
 Third, the simulator will compare the value of the time gap with the following scenarios (shown in Fig. \ref{ACCScenarios}):
 \begin{itemize}
 	\item \emph{TimeGap $>$ ($\Delta$TimeGap + safeTimeGap)}. This indicates that the vehicle ahead is so far from the point $t_{\sigma}$. The simulator will accelerate the speed of the vehicle robot till the desired speed. The simulator adjusts (increase$/$decrease) the current speed by using the following equation: 
 	\begin{equation} 
 	currentSpeed  +/-=  \sqrt   {speed^2 +  2 *  (Time) },  \end{equation}  where $Time = ((\Delta Timegap + safeTimeGap) - TimeGap)$  
 	\item \emph{(TimeGap $>$ safeTimeGap) $\&\&$ (timeGap $<$ ($\Delta$TimeGap  + safeTimeGap))}. This indicates that the vehicle robot ahead is approaching within the period of time gap between $[$$t_{\sigma}$  $t_{safeTimeGap}$]. The simulator will put the ACC system in \emph{follow} mode.  \emph{Follow} mode means that there is a vehicle in front in the lane. The simulator will automatically adjust the current speed by using equation 3.
 	\item \emph{TimeGap $==$ safeTimeGap}. This indicates that the vehicle robot ahead is approaching within the desired time gap and there is a safety distance between them. The simulator will put the ACC system in the cruise mode. \emph{Cruise} mode means that the vehicle robot ahead is approaching in safe time gap. Then, the simulator will set the current speed as the desired speed.  
 	\item \emph{TimeGap $< $ safeTimeGap}. This indicates that the vehicle ahead is moving within the time between $[$$t_{safeTimeGap}$ $t_{0}$$]$. The simulator will reduce the speed of the vehicle by using equation 3.   
 	
 	\item \emph{TimeGap $==$ 0}. This indicates that the vehicle ahead has fully stopped. Then the simulator will bring the vehicle to a complete stop at the standstill distance and change the ACC mode to stop. If the front vehicle starts to move again, then the simulator will change the ACC mode to resume. \emph{Resume} mode means that the current speed of the ACC vehicle will be accelerated to the desired speed. The simulator uses the following equation to achieve that: 
 	\begin{equation} 
 	currentSpeed  +=  accelerationratio,	\end{equation} \text{where  accelerationratio is set to 4 cm$/$sec;}

 \end{itemize}
 
 \subsection{Deriving Software Safety Requirements of the ACC Simulator}

 To derive the software safety requirements, we applied the STPA SwISs Step 1 to the system specification requirements. We used the XSTAMPP software tool to document the results of STPA and generate the formal specification of the STPA results. 
 
 As a result, we identified the system-level accidents that the simulator software can lead (or contribute to). For example, \textbf{ACC-1 : The ACC robot crashes the robot ahead}. The system-level hazards which can lead to this accident are:

 \begin{itemize}
 	\item $H_{1}$: The ACC software does not keep a safe distance from the a vehicle robot ahead.
 	\item $H_{2}$: The ACC software provides an unintended acceleration when the vehicle in front is too close.
 	\item $H_{3}$: The ACC software does not stop the vehicle when the vehicle ahead is fully stopped.
 \end{itemize}
 
 \begin{figure}[t]
 	\centering
 	\includegraphics[width=10cm]{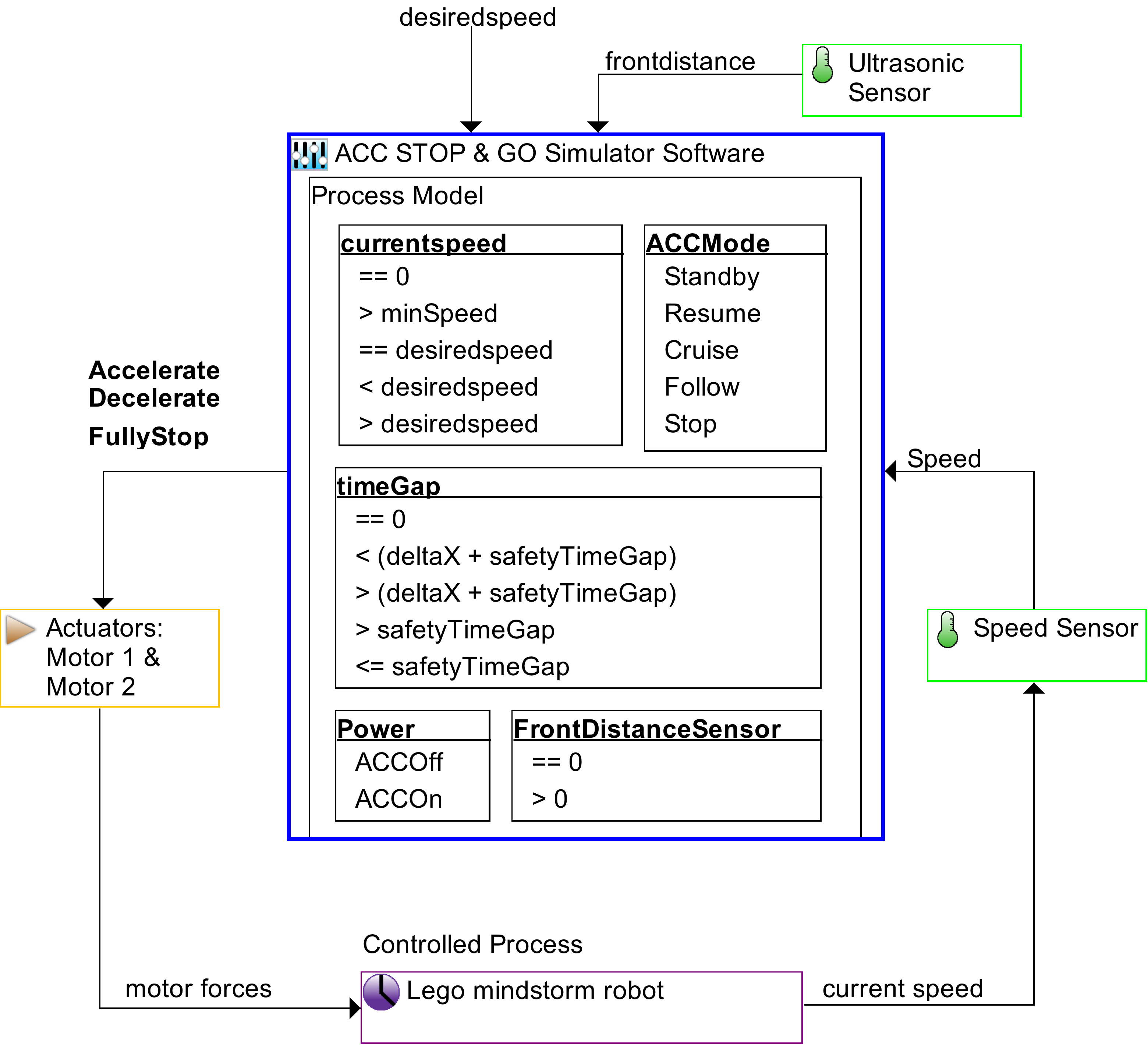}
 	\caption{The control structure diagram of ACC with the safety-critical process model variables}
 	\label{figACControlstructure}
 \end{figure}
 
 We built the control structure diagram of the ACC simulator (shown in Fig. \ref{figACControlstructure}). It contains the main interconnecting components of the ACC simulator at a high level, such as the \emph{ACC simulator software controller unit}, \emph{the electronic motors}, \emph{the robot vehicle} as the controlled process, and \emph{the Ultrasonic and speed sensors}. The ACC software controller receives the distance data from the ultrasonic sensor and current speed data from the speed sensor. Based on this information, the software will calculate the time gap and determine if the vehicle robot ahead is present. The ACC software will adjust the speed of the robot based on the above sensors and issues one of the critical safety control action: accelerate, decelerate, or fullystop. Each one of these control actions will be evaluated based on the four general hazardous types (columns of table II). Table II shows the examples of the potential unsafe control actions of the ACC simulator.

 \begin{table} [t]
 	
 	\def\arraystretch{1.1} 
 	\renewcommand{\arraystretch}{1.4}
 	
 	\tbl{Examples of potentially unsafe control actions of the ACC software controller} {
 		\begin{tabular}{|p{1.6cm}| p{2.2cm } |p{2.2cm }| p{2.2cm }| p{2.4cm }|}
 			\hline
 			{Control $~$Action} &  {Not providing causes hazard} &  {Providing causes hazard } &  {Wrong timing or order causes hazard } &  {Stopped too soon or Applied too long} \\
 			\hline
 			{Accelerate Speed} &  The ACC software does not accelerate the speed when the robot vehicle ahead is so far in the lane. [\textbf{Not Hazardous}] & \multicolumn{1}{ |p{3.2cm }|}{\textbf{UCA-1.1}: The ACC software accelerates the speed of robot unintendedly when the time gap to the robot vehicle ahead is smaller than the desired time gap. [\textbf{H-1}] [\textbf{H$-$2}]} & \textbf{UCA-1.2}: The ACC software accelerates the speed before the robot vehicle  ahead starts to move again. [\textbf{H-1}] [\textbf{H-2}] & \textbf{UCA-1.3}: The ACC software accelerates the speed too long so that it exceeds the desired speed of the robot. [\textbf{H-2}] \\
 			\hline
 			{Decelerate Speed} &  \textbf{UCA-1.5}: The ACC software does not decelerate the speed when the robot vehicle ahead is too close in the lane. [\textbf{H-1}] & \multicolumn{1}{ |p{3.2cm }|}{ The ACC software decelerates the speed of robot unintendedly when the time gap to the robot vehicle ahead is larger than desired time gap. [\textbf{Not Hazardous}]} & The ACC software decelerates the speed when the robot vehicle  ahead starts to move again. [\textbf{Not Hazardous}] & \textbf{UCA-1.6}: The ACC software decelerates the speed long enough so that it cannot bring the robot to fully stop when the robot ahead is stopped. [\textbf{H-3}] \\
 			\hline
 			{FullyStop} &   \textbf{UCA-1.4}: The ACC software does not bring the robot to a complete stop at a standstill when the robot vehicle ahead is fully stopped. [\textbf{H-1, H-3}] & \multicolumn{1}{ |p{3.2cm }|}{ The ACC software stops the robot suddenly when the distance to the robot ahead is too far. [\textbf{Not Hazardous}]} &  The ACC software does not accelerate the speed after the robot vehicle  ahead starts to move again. [\textbf{Not Hazardous}] & N/A \\
 			\hline
 			
 		\end{tabular} }

 	\end{table}%

 	 We evaluated each item in table II to check whether it can contribute or lead to any system-level hazards ($H_{1}-$ $H_{3}$). If an item is hazardous, then we assign one or more system-level hazards to it. We translate each hazardous item manually to the corresponding software safety requirement by using the guide words, e.g., \emph{have to, must be, or should}. Table III shows examples of the informal textual software safety requirements.

 	\begin{table}  
 		\def\arraystretch{1.1} 
 		\renewcommand{\arraystretch}{1.4}
 		\tbl{Examples of software safety requirements at the system level} {
 			
 			\begin{tabular}{ |p{1.2cm}| p{11.5cm}|}
 				\hline
 				\multicolumn{1}{ |p{2cm}| }{ {Related UCAs}} &  {Corresponding Safety Constraints} \\
 				\hline
 				\multicolumn{1}{|c|}{UCA-1.1 } & \textbf{SSR1.1}- ACC software must not accelerate the speed of the robot when the target robot vehicle is too close in the lane. \\	\hline
 				\multicolumn{1}{|c|}{UCA-1.2 } & \textbf{SSR1.2}- ACC software must not accelerate the speed when the robot ahead is fully stopped. \\	\hline
 				\multicolumn{1}{|c|}{UCA-1.3 } & \textbf{SSR1.3}-ACC software must not increase the speed  than the desired speed. \\	\hline
 				\multicolumn{1}{|c|}{UCA-1.4 } & \textbf{SSR1.4}-ACC controller must stop the robot at standstill point (shown in Fig. \ref{ACCScenarios}) when the robot ahead is fully stopped. \\	\hline

 			\end{tabular}}%

 		\end{table}
 		
 		\begin{table*}[t]
 			\def\arraystretch{1.1} 
 			\renewcommand{\arraystretch}{1.4}

 			\tbl{Examples of the context table of \emph{providing} the control action\emph{ accelerate} } {
 				
 				\begin{tabular}{ |p{1.1cm}|  p{2.5cm} | p{4.0cm} |  p{1.5cm} |  p{2.1cm}| }
 					\hline
 					\textbf{Control Actions} & \multicolumn{3}{|c| }{ {Process Model Variables}} &  { Is it a hazardous Control Action?} \\
 					\hline
 					\multicolumn{1}{ |p{2.2cm} |}{} &  {CurrentSpeed} &  {TimeGap} &  {ACC Mode} & \multicolumn{1}{|c| }{ {providing}} \\
 					\cline{2-5}\multicolumn{1}{|p{2.2cm} |}{\multirow{3}[6]{*}{\textbf{accelerate}}} &  CS$>$ minSpeed &  TimeGap $<$( $\Delta$ Timegap + safetyTimeGap) & follow & \multicolumn{1}{|c| }{No} \\
 					\cline{2-5}\multicolumn{1}{|c| }{} &  CS$<=$desiredSpeed  & TimeGap $==$ 0 & follow & \multicolumn{1}{|c| }{Yes, H2, H1} \\
 					\cline{2-5}\multicolumn{1}{|c| }{} &   CS$<$desiredSpeed & TimeGap $>$safetyTimeGap  & follow & \multicolumn{1}{|c| }{No} \\
 					\cline{2-5}\multicolumn{1}{|c| }{} &   CS$<$desiredSpeed &  TimeGap $<$ ($\Delta$ TimeGap + safetyTimeGap) & follow & \multicolumn{1}{|c| }{Yes} \\
 					
 					\hline
 				\end{tabular}}%

 			\end{table*}%
 			
 			\begin{table} 
 				\def\arraystretch{1.1} 
 				\renewcommand{\arraystretch}{1.4}
 				\tbl{Examples of unsafe software scenarios in XSTAMPP based on critical combinations} {
 					
 					\begin{tabular}{  p{1.0cm}  p{11.5cm} |  }
 						\hline
 						\multicolumn{1}{ |l|  }{ {ID}} &  { Unsafe software safety scenarios} \\
 						\hline
 						\multicolumn{1}{|c|}{RUCA-1.1 } & The ACC software controller provides the accelerate command when ACC mode is Standby and timeGap is greater than (deltaX + safetyTimeGap) and the current speed is less than desired speed. \\	\hline
 						\multicolumn{1}{|c|}{RUCA-1.2 } & The ACC software controller provides the accelerate command when timeGap is less than (deltaX +TimeGap). \ \\	\hline
 						\multicolumn{1}{|c|}{RUCA-1.3 } & The ACC software controller provides the accelerate command  when current speed is greater than or equal to desired speed. \\	\hline
 						\multicolumn{1}{|c|}{RUCA-1.4 } & \ The ACC software controller does not provide the fullyStop command when the timeGap is 0. \\	\hline
 						\multicolumn{1}{|c|}{RUCA2.1 } & The ACC software controller provides the decelerate command too late when ACC mode is follow and timeGap is less than safetyTimeGap and currentSpeed is greater than desired speed. 
 						\\
 						\hline
 					\end{tabular}}%
 					
 				\end{table}
 				
 				To refine the informal textual software safety requirements which are shown in table III, we identified the process model of the ACC software controller and its critical variables which have an effect on the safety of the ACC software control actions. Figure \ref{figACControlstructure} shows the control structure diagram and process model variables of the ACC software. The ACC software has three safety-critical process model variables: \emph{Internal variables} such as currentSpeed (5 values), Timegap (5 values), \emph{Internal states variable}  such as ACC mode (states) with 5 values, and \emph{the environmental variables} such as front distance. Each safety control action provided by the ACC software should be evaluated to determine whether it will be hazardous or not when the combination set of relevant values of the process model variables (context) occur. 
 				
 				We used XSTAMPP to generate the critical combinations (context tables) for each safety-critical action in the two contexts \emph{when the control action is provided} and \emph{it is not provided} and causes hazard. For each control action, the total number of combinations between the process model variables is (5 $\times$ 5 $\times$ 5  =125) combinations. We reduced the number of combinations by applying pairwise test coverage to the generated combination sets. The number of critical combinations is reduced to 25 for each control action. 
 				
 				Based on the generated combination sets, we evaluated each control action in two contexts \emph{Providing} and \emph{Not Providing}. Table IV shows examples of the context table of providing the control action \emph{accelerate} based on the combinations of the values of the critical process model variables. As a result, we identified 32 unsafe scenarios (shown in Table V) for all the control actions \emph{accelerate (18 scenarios)},  \emph{decelerate (7 scenarios )} and \emph{FullyStop (7 scenarios)}. Table VI shows the examples of generated software safety requriements for the unsafe scenarios.
 				
 				\begin{table} [t]
 					\def\arraystretch{1.1} 
 					\renewcommand{\arraystretch}{1.4}
 					\tbl{Examples of generated software safety requirements in XSTAMPP for the unsafe scenarios} {
 						
 						\begin{tabular}{  p{1.0cm}  p{11.5cm} |  }
 							\hline
 							\multicolumn{1}{ |l|  }{ {Related UCAs}} &  {Refined Safety Constraints} \\
 							\hline
 							\multicolumn{1}{|c|}{RUCA-1.1 } & \textbf{RSSR1.1}- Accelerate command must not be provided when ACC mode is Standby and timeGap is greater than (deltaX + safetyTimeGap) and the current speed is less than desired speed. \\	\hline
 							\multicolumn{1}{|c|}{RUCA-1.2 } & \textbf{RSSR1.2}- Accelerate command must not be provided when timeGap is less than (deltaX +TimeGap). \ \\	\hline
 							\multicolumn{1}{|c|}{RUCA-1.3 } & \textbf{RSSR1.3}-Accelerate command must not be provided when current speed is greater than or equal to desired speed. \\	\hline
 							\multicolumn{1}{|c|}{RUCA-1.4 } & \textbf{RSSR1.4}-FullyStop command must provided when the timeGap is 0. \\	\hline
 							\multicolumn{1}{|c|}{RUCA-2.1 } & \textbf{RSSR2.1}- Decelerate command must not be provided too late when ACC mode is follow and timeGap is less than safetyTimeGap and currentSpeed is greater than desired speed. 
 							\\
 							\hline
 						\end{tabular}}%
 						
 					\end{table}
 					
 \subsubsection{Formalizing the Software Safety Requirements of the ACC Software Simulator}					
 We formalised the STPA-generated software safety requirements of the ACC software simulator which are derived in Step 1 of the proposed approach (shown in Table VI). We used XSTAMPP to automatically refine the informal textual software safety requirements into formal textual software safety requirements (shown in Table VI). Based on the rules 3-4, XSTAMPP automatically generates the LTL formula for each refined software safety requirement. Table VII shows the examples of the corresponding LTL formula of each software safety requirement. We used the generated-LTL formulae to verify the safe behavioural model which is constructed from the STPA results.

 \begin{table} [t]
 	
 	\def\arraystretch{1.1} 
 	\renewcommand{\arraystretch}{1.4}
 	\tbl{Examples of LTL formulae of the refined software safety requirements at the system level} {
 		\begin{tabular}{  p{2.0cm}  p{11cm} | }
 			\hline
 			\multicolumn{1}{ |l|}{ {Refined SSRs}} &  {Corresponding LTL formula} \\
 			\hline
 			\multicolumn{1}{|c|}{RSSR1.1 } & \textbf{LTL1.1}-  G ((state=Standby) \&\& (timeGap $>$ deltaX+safetyTimeGap) \&\& (currenSpeed$<$desiredSpeed) \mbox{-$>$ !} (controlAction==Accelerate)). \\\hline
 			\multicolumn{1}{|c|}{RSSR1.2 } &  \textbf{LTL1.2}- G((currentSpeed $>$ desiredSpeed ) \&\& (TimeGap $<$(deltaTime + safetyTimeGap)) $-> !$ (controlAction==Accelerate). \\\hline
 			\multicolumn{1}{|c|}{RSSR1.3 } & \textbf{LTL1.3}-G((currentSpeed $>=$desiredSpeed) $-> !$ (ControlAction==stop) . \\\hline
 			\multicolumn{1}{|c|}{RSSR1.4 } &\textbf{LTL1.4}- G((timeGap $==0$) $->$X  (controlAction==FullyStop). \\\hline
 			\multicolumn{1}{|c|}{RSSR2.1 } & \textbf{LTL.2.1}- G ((state==Follow) \&\& (timeGap $<$safetyTimeGap) \&\& (currentSpeed $>=$ desiredSpeed) $->$ !(controlAction==Decelerate))	\\ 
 			\hline
 		\end{tabular}}%

 	\end{table}
 	
 \subsection{Automatically Generating SMV Model}
 
\begin{figure*}[t]
 	\centering
 	\includegraphics[width=5.5in]{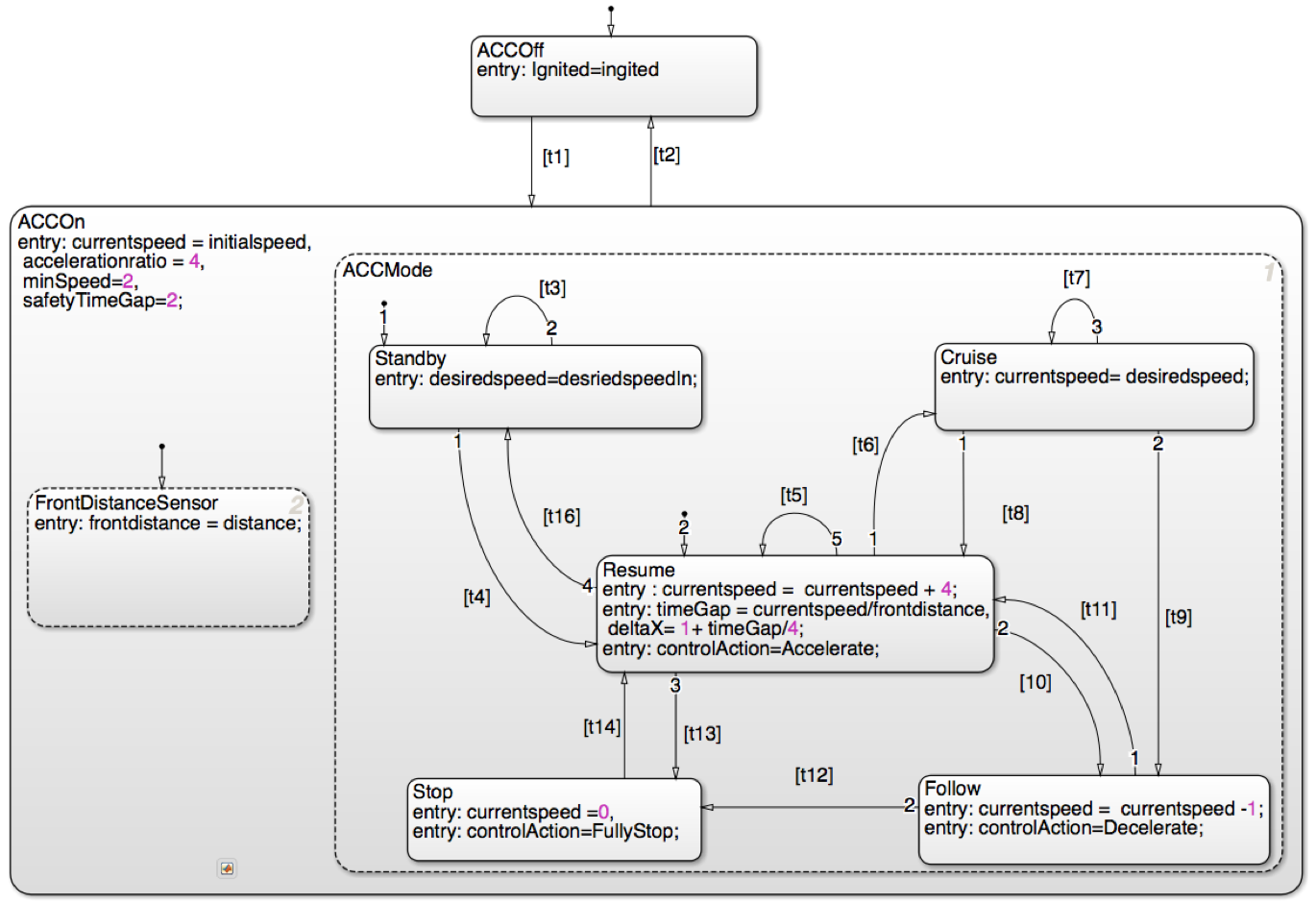}
 	\caption{The safe behavioural model of the ACC software controller}
 	\label{safebehavoiral}
 \end{figure*}

 We visualised the process model of the ACC software controller (shown in Fig. \ref{figACControlstructure}) by createding a Simulink/Matlab Stateflow model (shown in Fig. \ref{safebehavoiral}). The Stateflow contains 9 states (2 of them are superstates) and 19 transitions. It shows the relationship between the process model variables in the safety control structure diagram of the ACC simulator. The process model describes the critical variables and states of the software and how the software issues the critical safety control actions (e.g. accelerate, decelerate, etc.)
 
   We generated the SMV model of the safe behavioural model (shown in Fig. \ref{figACControlstructure}) by using the STPA TCGenerator tool which transforms the safe behvioural model into a verification input of the NuSVM model checker. For that, we first derived the XML specifications of the Simulink's Stateflow model. Second, we took the XML specifications of both ACC simulator STPA file and the safe behavioural model as input to the STPA TCGenerator tool. The tool parses both files and generates the SMV model which maps all states, transitions and data variables, and LTL formulae of STPA software safety requirements of the safe behavioural model to SMV model specifications.
 
 We updated the default values of each input data variable which are declared in the generated SMV model (e.g. \emph{initial speed (10.0), desired speed (45.0), initial frontdistance (150.0)}). The value of current speed will be calculated by using equations 5. The value of time gap will also be calculated by using equation 2. The STPA TCGenerator tool runs the NuSMV 2.6.0 model checking tool to verify the generated SMV model file. The NuSMV model succeeded in verifying the generated SMV model within 0.29 seconds and no further errors were reported. NuSMV consumed 42.10 megabytes to store 2.31828e+17 states and performed 2.97418e+09 transitions. As a result, all LTL formulae were satisfied and there is no counterexample generated because the safe behavioural model itself was built from STPA software safety requirements. 
 
 \subsection{Safety-based Test Case Generation from the Safe Test Model}
 After validating the correctness of the safe behavioural model, we used the STPA TCGenerator to generate a hierarchical tree of the safe behavioural model which shows the hierarchy levels of the safe test model. The \emph{STPA TCGenerator} tool parses the tree of the safe behavioural model recursively by considering superstate decompositions \emph {AND\_STATE} (parallel) and \emph{OR\_STATE} (exclusive) to generate the safe test model as an extended finite state machine. As a result, the generated safe test model contains 7 states (after removing the superstates) and 32 transitions (after maintaining the transitions of superstates). The tool automatically generates the traceability matrix between STPA software safety requirements and the safe behavioural model. 
 
 To generate the safety-based test cases from the safe test model of the ACC simulator, we first set the number of test steps to 10 and selected the three test coverage criteria (state, transition and STPA software safety requirements test coverage criteria) in the STPA TCGenerator tool. We selected the STPA software safety requirements coverage as the stop condition of the test case generating algorithm. We also set the test input value for each input data variable: \emph{power} (true), desired speed (45 cm/sec), initial speed (10 cm/sec), front distance (150 cm). Finally, we ran the STPA TCGenerator tool three times to generate safety-based test cases from the test model, respectively: 1) depth-first search, 2) breadth-first search and 3) the combined algorithm. Table VIII shows the results of the generated safety-based test cases by each test algorithm. We could achieve 100\% coverage of all the STPA software safe requirements which are linked to the safe test model in the traceability matrix. Figure \ref{testcase} shows an example of the format of documenting each safety-based test case.
 
 \begin{figure} [h!]
 	
 	\begin{minipage}{8,7cm}
 		\begin{lstlisting}[language=Python, numbers=left ]
 [Test Case ID] 2  
 [Test Suite ID] 2
 [Related STPA SSRs]
   RSSR1.1, RSSR1.2, RSSR1.3
 [PreConditons]
   desiredspeed=45.0
   frontdistance=120.32
   currentspeed=44.0
   state=Resume
 [Actions]
   controlAction=Accelerate
 [PostConditons]
   currentSpeed=45.0
   state=Cruise
 [Comment]
 		
 		
 		
 		
 		
 		\end{lstlisting}

 	\end{minipage}
 	\caption{An example of a generated safety-based test case}
 	\label{testcase}
 \end{figure} 
 
 \begin{table} [t]%
 	\def\arraystretch{1.1} 
 	\renewcommand{\arraystretch}{1.4}
 	\tbl{The safety-based test cases generated by STPA TCGenerator tool \label{tab:one}}{%
 		\begin{tabular}{|p{0.4cm}|p{1.2cm}|p{0.8cm}|p{0.8cm}|p{0.8cm}|p{1.1cm}|p{1.6cm}|p{1.8cm}| p{1.6cm}|  }
 			\hline
 			{	ID }&  {Test Algorithm }&  {Test Steps} &  {Test Suite} &  {Test Cases }&  {Time (in Sec) }&  {State ~~~Coverage} &  {Transition Coverage} &  {STPA SRR Coverage }\\ \hline
 			1  & DFS            & 10         & 1          & 119                          & 3            & 6/7 = 85.7\%    & 23/32=71.9\%       & 32/32=100\%     \\  \hline
 			2  & BFS            & 10         & 4         & 24                               & 1             & 6/7 = 85.7\%    & 17/32= 53.1\%      & 32/32=100\%     \\  \hline
 			3  & Both           & 10         & 5        & 249                          & 2            & 7/7 = 100\%    & 18/32= 87.5\%      & 32/32=100\% \\ \hline    
 		\end{tabular}}
 		
 	\end{table}%

\begin{figure}[t]
 		\centering
 		\includegraphics[width=5.5in]{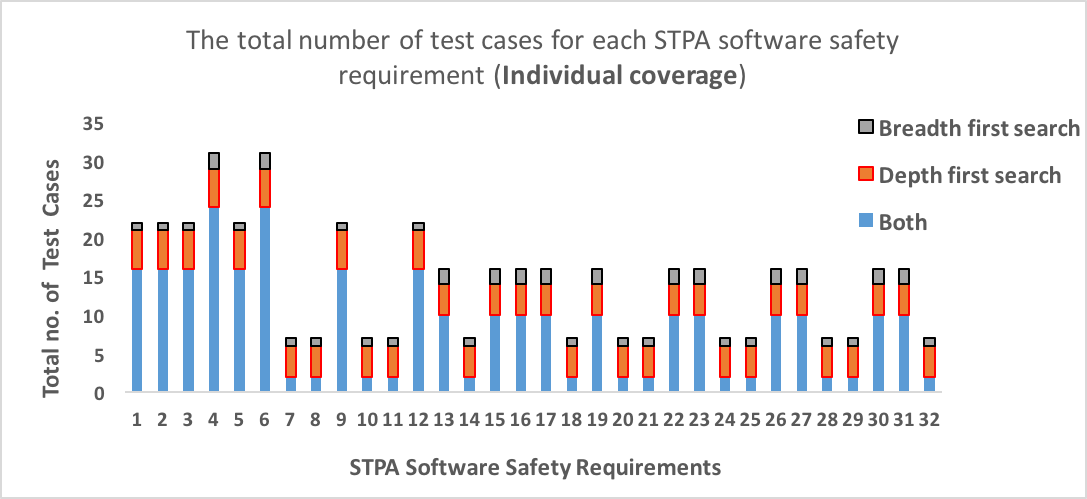}
 		\caption{The total number of test cases for each STPA software safety requirement}
 		\label{totalnumber}
 	\end{figure}

 	Based on the traceability matrix between the model and the STPA software safety requirements, the \emph{STPA TCGenerator} provides an \emph{ individual coverage} (how many test cases $TC$ covered each $SSR$)  by each test algorithm (shown in Fig. \ref{totalnumber}). 
 	
\section{Discussion}

The idea behind the proposed approach is to integrate STPA safety analysis and its identification of the hazardous situations that the software can lead or contribute to semi-automatically with software testing. For this, we formalise the STPA software safety requirements into a formal specification and model the information derived from the STPA safety analysis into a test model.  That helps us to focus the effort of testing by generating safety-based test cases for each software safety requirements. However, there are still some open issues and interesting challenges that require further research.  

\subsection{Visualisation of  Process Model}
The process model in the STPA control structure diagram is a very abstract model which shows only the safety-critical variables and states which have an effect on the safety of issuing the control actions by a software controller in the control structure diagram. It does not show how the software controller issues the control actions. Therefore, we use the statechart notation to visualise the relationships of the process model variables and describe the safe behavioural model. However, constructing a safe behavioural model from the STPA safety results by safety test engineers depends on the level of information which is available during the STPA safety analysis process (e.g. process model, and process model variables and values). Moreover, it is critical how this information describes the internal state of the software controller and the safety-critical software variables (e.g.\ interaction and environmental variables). Furthermore, visualising the safe behavioural model in a modelling tool such as Simulink requires user expertise in the modelling of dynamic behaviour to map the safety analysis specifications (process model, control actions and software safety constraints) into the Stateflow notation. Therefore, this point remains as future work to automatically provide a basic structure of the safe behavioural model from the process model information (e.g.\ states and its hierarchical levels) which is visualised in XSTAMPP. This will help the safety tester to understand the relationships between the critical system states, environmental and interaction variables which are documented in the process model of the software controller in the STPA control structure diagram. 

\subsection{The Correctness of the Safe Test Model}
The manual construction of test cases is a hard, time-consuming and error-prone activity that requires deep knowledge and expertise. Furthermore, the manual building of a test model from system specifications with the purpose of generating test cases still needs a proof of its correctness to ensure that the test model captures all specifications. A solution is to construct a test model for a given system and prove its correctness by transforming it into an intermediate model which is supported by a formal verification approach (e.g. model checker) to verify the generated model against its specifications. In addition, the specifications should also be mapped from informal text to the formal specifications. For this issue, we transformed the safe test model into the SMV model and verified it by using the NuSMV model checker to ensure that the safe test model satisfies the STPA specifications. However, the model transformation process also needs a proof of the correctness of the resultant model, even though the model checker did not induct any error. In our proposed approach, this issue remains as an open issue for future work. 

\subsection{Traceability Matrix}
The automation of the test case generation process can lead to a large number of test cases that cover the same information. Reducing the number of generated test cases is a major factor in evaluating the effectiveness of an automated testing tool and the quality of the generated test cases. Therefore, we added a new test coverage criteria (STPA software safety requirements) to stop the test case generating algorithm when this criterion becomes 100\% to ensure that each STPA safety requirement is covered at least in one test case. Furthermore, the first prototype of the \emph{STPA TCGenerator} tool supports to generate test cases for each software safety requirement by automatically generating a traceability matrix by calculating  the similarity degree of the matched tokens between the STPA software safety requirements and the safe test model. The traceability matrix contains all relevant transitions of each software safety requirement in the safe test model. 

\subsection{Process Model Variables Data Types}
Another limitation is that the process model variables in the STPA control structure diagram visualised by XSTAMPP have no data types. Furthermore, XSTAMPP does not support multi-levels hierarchies of the process model of the software controller in the control structures. That makes ensuring and checking the consistency between the hierarchy levels of the process model in STPA and the Stateflow model in Simulink a big challenge. For example, the process model variable \emph{ACC} Active in the \emph{ACC} software controller has sub-process model variables such as control speed and \emph{FrontDistancesensor}  which will be activated when the ACC state is active. Therefore, it requires human effort to define the process model hierarchy and map it to the Simulink Stateflow model hierarchy level.

\section{Conclusion}

In this paper, we introduced a systematic and semi-automatic approach to generate safety-based test cases based on the STPA safety analysis. Our approach concentrates on generating a set of test cases for each STPA software safety requirement. The generated test cases will be used to verify the safety of the software-intensive system under analysis. We also implemented an open-source tool support that automates the safety-based test cases generating approach. Furthermore, we illustrated the proposed approach with safety-critical software of an ACC system with stop-and-go function. The results show that deriving test cases based on the safety requirements is a practical and effective approach to generate different test cases to recognize software risks and assure the software quality.  

As a future work, there are many interesting directions and trends to extend the research of safety-based testing for software-intensive systems and the automated tool support. We plan to improve the tool by considering the other Stateflow semantics which were not addressed in our approach such as inner transitions and connective and history junctions. Furthermore, we aim to limit the number of the generated test cases, to improve the traceability matrix by adding information about the maximum number of test cases for each software safety requirement and also the priority value to generate a reasonable test case for each software safety requirement.

 Furthermore, we plan to improve the process model in the control structure diagram by allowing the safety analyst to define the data type of each process model variable and draw the multi-hierarchy levels of the process model variables. Finally, we plan to evaluate the proposed approach and the tool support on a real software-intensive system with an industrial partner.

\begin{acks}
 The authors would like to thank Prof. Nancy Leveson, MIT, for her very careful review of our paper, and for the comments, corrections and suggestions that ensued. 
 
 We would also like to express gratitude to Lukas Balzer, University of Stuttgart,  who worked with us to improve and build the XSTAMPP platform;  Yannic Sowoidnich, University of Stuttgart, who developed XSTPA; Ting Luk-He, University of Stuttgart, who developed an Eclipse plugin for STPA TCGenerator; Rick Kuhn, National Institute of Standards and Technology, USA, who provides us the Automated Combinatorial Testing Tool (ACTS). We are grateful for their help, effort and time; and  Kornelia Kuhle, University of Stuttgart, for her feedback on the text.  We also would like to express our appreciation to the anonymous referees for their in-depth comments, suggestions and corrections to improve the quality of the paper.

\end{acks}


  \bibliographystyle{ACM-Reference-Format-Journals}
 \bibliography{acmsmall-sample-bibfile}


\end{document}